\documentclass[pra,groupedaddress,superscriptaddress,onecolumn,amsmath,amssymb,a4paper,preprint,notitlepage]{revtex4}
\usepackage{geometry}
\usepackage[dvipsnames]{xcolor}
\usepackage{graphicx}
\usepackage{verbatim}
\usepackage{float}
\usepackage{graphics}
\usepackage{mathrsfs}
\usepackage{amssymb}
\usepackage{amsmath}
\usepackage{amsthm}
\usepackage{bm}
\usepackage{dsfont}
\usepackage{hyperref}
\usepackage{braket}
\usepackage{wrapfig}
\usepackage{amsbsy}
\usepackage{color}
\usepackage[normalem]{ulem}
\usepackage{subfig}

\usepackage[T1]{fontenc}

\newcommand{\be}{\begin{equation}}
\newcommand{\ee}{\end{equation}}
\newcommand{\ba}{\begin{eqnarray}}
\newcommand{\ea}{\end{eqnarray}}
\newcommand{\bibt}{\bibitem}

\begin{document}

\title{Evolution of a Non-Hermitian Quantum Single-Molecule Junction at Constant Temperature}

\author{Andrea Grimaldi}

\affiliation{Dipartimento di Scienze Matematiche e Informatiche,
Scienze Fisiche e Scienze della Terra, Universit\`a degli
Studi di Messina, viale F. Stagno d’Alcontres 31, 98166 Messina,
Italy}

\author{Alessandro Sergi} 

\affiliation{Dipartimento di Scienze Matematiche e Informatiche,
Scienze Fisiche e Scienze della Terra, Universit\`a degli
Studi di Messina, viale F. Stagno d’Alcontres 31, 98166 Messina,
Italy}

\affiliation{Istituto Nazionale di Fisica Nucleare,
Sez. di Catania, 95123 Catania, Italy}

\affiliation{Institute of Systems Science, Durban University of Technology,
P. O. Box 1334, Durban 4000, South Africa}

\author{Antonino Messina}

\affiliation{Dipartimento di Matematica ed Informatica dell’Universit\`a di Palermo,
Via Archirafi 34, I-90123 Palermo, Italy; antonino.messina@unipa.it}

\begin{abstract}
This work 
concerns the theoretical description of the quantum dynamics of molecular junctions with
thermal fluctuations and probability losses
To this end, we propose a theory for describing non-Hermitian quantum 
systems embedded in constant-temperature environments.
Along the lines discussed in
[A. Sergi \emph{et al}, \emph{Symmetry} {\bf 10} 518 (2018)],
we adopt the operator-valued Wigner formulation of quantum mechanics
(wherein the density matrix depends on the points of the Wigner phase space associated to the system) and derive a non-linear equation of motion.
Moreover, we introduce a model  for
a non-Hermitian quantum single-molecule junction (nHQSMJ).
In this model the leads are mapped to a tunneling two-level system,
which is in turn coupled to a harmonic mode (\emph{i.e.}, the molecule).
A decay operator acting on the two-level system describes
phenomenologically probability losses.
Finally, the temperature of the molecule is controlled by means
of a Nos\'e-Hoover chain thermostat.
A numerical study of the quantum dynamics of this toy model 
at different temperatures is reported.
We find that the combined action of  probability losses and thermal fluctuations
assists quantum transport through the molecular junction.
The possibility that the formalism here presented can be extended
to treat both more quantum states ($\sim 10$) and many more
classical modes or atomic particles ($\sim 10^3-10^5$) is highlighted.
\end{abstract}

\maketitle




%%%%%%%%%%%%%%%%%%%%%%%%%%%%%%%%%%%%%%%%%%

\section{Introduction}

Molecular junctions are nano-devices composed of metal or
semiconductor electrodes known as leads \cite{hystory,thoss}.
A single molecule simulates a conducting bridge between the leads.
Such a nanostructure is almost perfectly suited to study
non-equilibrium quantum transport \cite{hystory,thoss}.

Various approaches satisfactorily befit the investigation of the phenomenology
of molecular junctions \cite{thoss}.
In this work, we introduce a simple toy model of a molecular junction
\cite{thoss} in order to perform a qualitative study aimed at singling out
universal features of such systems. To this end, the operator-valued Wigner formulation
of quantum mechanics \cite{quasi_lie,zhang-balescu,balescu-zhang} is
particularly useful to embed quantum toy-models in quantum phase space baths
since it can simultaneously describe dissipative phenomena and thermal fluctuations.
A quantum-classical molecular junction model  has been studied in Ref. \cite{hanna2}.

%%%%%%% non-Hermitian Hamiltonians %%%%%%%%%%%%%%%%%%%%%%%%%%%%%%%%

In general, theories of quantum  transport offer an acceptable representation
of excitations and state population transfer.
They usually assume that what is transferred does not
disintegrate or disappear from the system. Such a physical property
is expressed through the conservation of probability.
On the contrary, by definition, the probability in quantum systems
with gain and loss is not conserved. Such a feature can be incorporated
in the theory by introducing non-Hermitian Hamiltonians
\cite{gamow_1928,carlbenderpt,mostafazadeh_2002,
subotnik,zelovich,zelovich2}.
Recently, the development of non-Hermitian quantum mechanics (NHQM)
has been propelled by various experiments on many systems 
(see \cite{rotter_15,moiseyev_11} and references therein).
While in this work a non-equilibrium quantum statistical mechanics of molecular junctions
is adopted, linear equations of motion are used  in Refs.  \cite{subotnik,zelovich,zelovich2}.
Conceptually, this can be understood in terms of the different theoretical targets:
In Refs.  \cite{subotnik,zelovich,zelovich2} the target is to study
non-equilibrium transport \emph{per se}. Instead, our work is aimed at showing the
possibility of a statistical mechanical description for non-Hermitian
quantum systems in classical bath \cite{as_15}.

Since probability non-conserving quantum systems are 
open quantum systems \cite{rotter_15}, it is natural to formulate NHQM 
in terms of the density matrix
\cite{as_15,sz_13,sz_15,sz_16,entropy2,grima_18,nh_emwave,sust}.
Typically, a non-Hermitian Hamiltonian is either derived by
means of the Feshbach formalism \cite{f-58,f-62} or it is postulated
as an \emph{ansatz} 
(see Refs.
\cite{sz_13,as_15,sz_15,sz_16,entropy2,grima_18,nh_emwave,sust}).
It is remarkable that there is a third possibility according to which a 
non-Hermitian Hamiltonian may arise from stroboscopic measurements performed
on an ancilla-subsystem added to a quantum system, 
a very interesting process capable of generating entanglement \cite{noi}. 
Given a non-Hermitian Hamiltonian and assuming that the Schr\"odinger equation
is still valid, one derives a probability non-conserving equation of motion
for the density matrix of the system \cite{noi}.
However, in order to meet the need of establishing a proper
quantum statistical theory, this non-Hermitian equation of motion
for the density matrix is generalized into a non linear-one
\cite{as_15,sz_13,sz_15,sz_16,entropy2,grima_18,nh_emwave,sust}.
Essentially, this is the origin of the difference between the linear approach of
\cite{subotnik,zelovich,zelovich2}
and the non-linear approach of
\cite{as_15,sz_13,sz_15,sz_16,entropy2,grima_18,nh_emwave,sust}.

Various investigations on thermal effects in molecular junctions are found in the literature
\cite{jtherm3,jtherm4,jtherm5}.
However, to the best of our knowledge, thermal effects have been investigated
separately from probability losses of molecular junctions.
In light of the previous considerations, the purpose of this work is
to combine the non-linear equation for the density matrix of a system with
a non-Hermitian Hamiltonian and temperature control in quantum phase space dynamics \cite{ap}.
Temperature control is technically implemented by means
of the Nos\'e-Hoover chain \cite{nhc} thermostat, formulated in terms of
quasi-Lie brackets \cite{quasi_lie}.
The unified formalism here reported is developed for studying
a greater variety of phenomena
than those accessible to theories treating thermal fluctuations
 and probability losses separately.
In detail, the theory is obtained through the density matrix approach
to non-Hermitian quantum mechanics
\cite{as_15,sz_13,sz_15,sz_16,entropy2,grima_18,nh_emwave,sust}
and it is expressed in terms of the operator-valued formulation of quantum mechanics
\cite{quasi_lie,zhang-balescu,balescu-zhang,as_15}.
For clarity, we highlight the three mathematical ingredients that must be
combined for building our formulation.
First, we need an approach for describing quantum operators in terms of
operator-valued Wigner phase space functions.
The conceptual framework of this specific  theory  assumes an 
Eulerian picture where a Hilbert space, or a space of quantum discrete states,
is defined for each point $X=(Q,P)$ of phase space. 
Accordingly, the operator-valued Wigner function \cite{quasi_lie,zhang-balescu,balescu-zhang}
may also be called phase-space-dependent density matrix.
The consistent coupling of quantum and phase space degrees of
freedom requires that quantum dynamics in the Hilbert space associated to $X$ also
depends non-locally on the quantum dynamics associated to different points $X'$.
Likewise, the dynamics of $X$ does not only depend on its quantum space
but it also depends on the quantum spaces of other points $X'$ and \emph{viceversa}. 
In practice, such a dependence can be implemented through a 
variety of algorithms \cite{theochem,sstp,algo12,algo15}.
When the Hamiltonian is quadratic, one can 
expand the operator-valued Moyal bracket \cite{b3,b4} up to linear order
in $\hbar$ and obtain quantum equations of motion \cite{mackernan}.

The second ingredient entering our theory is the
requirement that the degrees of freedom in phase space
evolve in time satisfying the constraint determined by the constant thermodynamic temperature.
The temperature control of the classical degrees of freedom is achieved \emph{in silico}
by means of the Nos\'e-Hoover chain algorithm \cite{nhc}.
We find such an approach advantageous because the formulation
of the Nos\'e-Hoover chain algorithm can be realized within the theoretical framework given by
the quasi-Lie brackets \cite{quasi_lie}. Such brackets allow one
to generalize the Nos\'e-Hoover chain algorithm, originally formulated for
classical systems only, to the more general case of the operator-valued
formulation of quantum mechanics  \cite{ap,b3,b4}.

The third ingredient, of course, is that the modeling of the leads 
provides a non-Hermitian quantum system \emph{per se}.

When all these three theoretical features are simultaneously taken into account, we obtain an equation that generalizes
the one arising from the quantum-classical quasi-Lie bracket \cite{quasi_lie} because it can describe probability gain/loss \cite{as_15} too.
Moreover, since such an equation keeps
 the thermodynamic temperature of the phase space degrees of freedom 
constant by means of the Nos\'e-Hoover chain algorithm \cite{quasi_lie,nhc,b3,b4},
it generalizes the equations used in  Refs.
\cite{as_15,sz_13,sz_15,sz_16,entropy2,grima_18,nh_emwave,sust},
which did not implement any thermodynamic constraint.

In this paper we investigate the time evolution 
of a single-molecule junction toy-model represented by a 
non-Hermitian Hamiltonian with a parametric phase space dependence.
The isolated leads are represented by a two-level system. 
In this situation, the transport of the occupation probability
of one or the other state can only occur through quantum tunneling.
The molecule connecting the two leads is modeled by a harmonic oscillator. 
When we consider only the coupling between the harmonic mode 
and the two-level system, the total system is considered closed.
In this case, transport can take place both \emph{via} tunneling
and \emph{via} the channel given by the oscillator.
In order to build a complete non-Hermitian quantum single-molecule junction
model at constant temperature,
the closed spin-boson system is transformed into an open quantum system
by means of two theoretical steps. The first step is to consider a
decay operator acting on the two-level system.
This operator introduces another transport channel that does not
conserve probability and then breaks time-reversibility.

The second step is to put the harmonic mode into contact with a heat bath. 
Of course, this condition forces the molecule to have the same
temperature of the bath and to experience fluctuations
according to the canonical ensemble probability distribution.
In practice, the action of the constant-temperature bath on
the molecule is implemented by means of the Nos\'e-Hoover chain thermostat.
For this reason, in the following
we will use wording such as ``heat bath'', ``constant-temperature
environment'' or ``Nos\'e-Hoover chain thermostat'' interchangeably.

The final complete non-Hermitian Hamiltonian is designed to represent
two lossy leads interacting with a thermalized molecule.

%%%%%%%%%%%%%%%%%%%%%%%%%%%%%

The organization of the paper is the following:
In Sec. \ref{sec:qdnhh} we give a short summary of the density matrix approach
to the dynamics of a system governed by a non-Hermitian Hamiltonian.
In Sec. \ref{sec:nhscb} we describe briefly how to treat 
non-Hermitian quantum systems  interacting with classical degrees of freedom.
We present the main theoretical results of this work in Sec. \ref{sec:nhscltb},
where we introduce the mathematical formalism
describing the quantum dynamics of a non-Hermitian quantum system interacting 
with classical degrees of freedom at constant temperature.
In Sec. \ref{sec:dmjtb} we introduce the non-Hermitian quantum single-molecule junction model.
The results of the numerical calculations are reported in Sec. \ref{sec:results}.
Finally, our conclusions are given in Sec. \ref{sec:conclusion}.

%%%%%%%%%%%%%%%%%%%%%%%%%%%%%%%%%%%%%%%%%%%

\section{Quantum dynamics of non-Hermitian systems}
\label{sec:qdnhh}

Consider a quantum system described by a non-Hermitian Hamiltonian operator
\be
\hat{\cal H} = \hat H - i \hat \Gamma \;,
\label{eq:defcalH}
\ee
where $\hat H$ and $\hat \Gamma$ are Hermitian operators
($\hat \Gamma$ is the decay operator).
In the presence of a probability sink or source, the dynamics of the
system is defined in terms of the following equations 
(we set $\hbar=1$):
\ba
\frac{d}{dt}|\Psi(t)\rangle &=& -i\hat{\cal H}|\Psi(t)\rangle
\;,
\label{eq:lsceq}
\\
\frac{d}{dt}\langle\Psi(t)|&=&i\langle\Psi(t)|\hat{\cal H}^\dag
\;,
\label{eq:rsceq}
\ea
where $|\Psi(t)\rangle$ and $\langle\Psi(t)|$ are state vectors of the system.
Equations (\ref{eq:lsceq}) and (\ref{eq:rsceq}) reduce to the standard
Schr\"odinger equation when $\hat \Gamma =0$.

The density matrix approach to non-Hermitian quantum systems
\cite{as_15,sz_13,sz_15,sz_16,entropy2,grima_18,nh_emwave,sust}
is obtained
upon introducing at any t>0 a Hermitian, semipositive defined, but 
non-normalized density matrix
\be
\hat \Omega(t) = |\Psi(t)\rangle\langle\Psi(t)| \;.
\label{eq:defOm}
\ee
The equation of motion for $\hat{\Omega}$ \cite{sz_13,grima_18,sust}
is obtained by taking the time derivative of Eq. (\ref{eq:defOm})
and using Eq. (\ref{eq:defcalH}):
\begin{equation}
\frac{d}{dt}\hat{\Omega}(t)=-i\left[\hat{H},\hat{\Omega}(t)\right]
-\left[ \hat{\Gamma},\hat{\Omega}(t)\right]_+ \;,
\label{eq:ddt-Omega}
\end{equation}
where $[...,...]$ is the commutator and $[...,...]_+$ is the anticommutator.
Equation (\ref{eq:ddt-Omega}) is linear but, nevertheless, it
describes an open quantum system. This is possible since it is defined
in terms of a decay operator that physically describes the presence
of probability sources/sinks, arising from the hidden
system with which the open quantum system interacts.

Equation (\ref{eq:ddt-Omega}) is non-Hermitian, breaks time reversibility and
does not conserve the trace of $\hat{\Omega}(t)$.
In fact, from Eq. (\ref{eq:ddt-Omega}) one can easily derive
\be
\frac{d}{dt}{\rm Tr}\left(\hat \Omega(t)\right)=
-2{\rm Tr}\left(\hat \Gamma\hat \Omega(t)\right)
\;.
\ee

In order to obtain a well-founded statistical mechanics of non-Hermitian quantum systems,
one can introduce the normalized density matrix
\be
\hat \rho(t)=\frac{\hat\Omega(t)}{{\rm Tr}\left(\hat \Omega(t)\right)}
\;.
\label{eq:defrho}
\ee
This operator $\hat\rho(t)$ obeys the non-linear equation
\begin{equation}
\frac{d}{dt}\hat\rho(t)=
-i\left[\hat H,\hat\rho(t)\right]
-\left[\hat\Gamma,\hat\rho(t)\right]_+
+2\hat{\rho}(t){\rm Tr}\left(\hat\Gamma\hat\rho(t)\right)
\;.
\label{eq:ddt-rho_matrix}
\end{equation}
As a consequence of its very definition and
of Eq. (\ref{eq:ddt-rho_matrix}), ${\rm Tr}(\hat\rho(t))=1$ at all times.
Equation (\ref{eq:ddt-rho_matrix}) breaks time reversal symmetry
and is non-linear.
In practice, the formalism trades off the non-conservation of probability in
Eq. (\ref{eq:ddt-Omega})
with the non-linearity of Eq. (\ref{eq:ddt-rho_matrix}).
As the linear Eq. (\ref{eq:ddt-Omega}), the non-linear
Eq. (\ref{eq:ddt-rho_matrix}) describes the dynamics of
an open quantum system.
Consistently, averages of arbitrary operators, \emph{e.g.}, $\hat{\chi}$,
can be calculated by means of the normalized density matrix $\hat{\rho}$ as
\be
\langle\hat{\chi}\rangle_{(t)}
={\rm Tr}\left(\hat\rho(t)\hat\chi \right)\;.
\label{eq:avenhs}
\ee
Equation (\ref{eq:avenhs}) establishes the foundations of the
statistical interpretation of non-Hermitian quantum mechanics.

\section{Non-Hermitian quantum systems set in phase space}
\label{sec:nhscb}

Non-Hermitian quantum systems can be embedded in phase space \cite{as_15}.
In the following, we sketch the derivation given in \cite{as_15}.
Let us consider a system described by a hybrid set of coordinates
$(\hat x,X)$, where $\hat x$ are quantum coordinates (operators)
and $X=(Q,P)$ are phase space coordinates
(an obvious multidimensional notation is used in order not to clutter
formulas with too many indices).
In this specific case, the $X$s describe the phase space coordinates
of the system.
As we have already discussed in the Introduction, the description of the system,
using the operator-valued-formulation of quantum mechanics,
can be established \emph{via} an Eulerian point of view
according to which a space of state vectors is defined
 for any point $X$ of phase space, 
\cite{zhang-balescu,balescu-zhang,quasi_lie}
so that $\hat x$ (and operators defined in terms of $\hat x$) can act on it.
In such a hybrid space 
there are operators that depend only on $\hat x$, which we denote
with the symbol $\hat{\phantom{x}}$ on top of the operator, 
\emph{e.g.}, $\hat\Gamma$.
There are pure phase space functions which we denote by their arguments,
\emph{e.g.}, $H_{\rm B}(X)$, or $G(X,t)$ when there is also an explicit time
dependence. Finally, there are operators that depend
both on $\hat x$ and (parametrically) on  $X$.
We denote the latters with the symbol $\tilde{\phantom{x}}$ on top, 
\emph{e.g.}, $\tilde\Omega(t) \equiv\hat\Omega(X,t)$.

In the non-Hermitian case, we consider a Hamiltonian operator
with parametric dependence on phase space
\ba
\tilde{\cal H}&=&\hat H + H_{\rm B}(X)+ \tilde H_I -i\hat\Gamma
\nonumber\\
&=& \tilde H_{\rm S} -i\hat\Gamma \;,
\label{eq:defcalHnh}
\ea
where $\hat H$ is the Hermitian Hamiltonian of the quantum subsystem,
$H_{\rm B}(X)$ is the Hamiltonian of the degrees of freedom in phase space,
$\tilde H_I$ describes
the interaction between the quantum system and the bath,
and $\tilde H_{\rm S}=\hat H + H_{\rm B}(X)+ \tilde H_I$.
Of course, $\tilde{\cal H}$ is the total non-Hermitian Hamiltonian
of the system with $\hat\Gamma$ as the decay operator.

When the operator $\hat\Gamma$ acts on the quantum dynamical variables
of the non-Hermitian quantum system, it has been shown~\cite{as_15} that,
as a generalization of the equations given in Refs. 
\cite{zhang-balescu,balescu-zhang,quasi_lie},
the equation of motion becomes
\begin{equation}
\frac{\partial}{\partial t}\tilde\Omega(t)=
-i\left[\tilde{H}_{\rm S},\tilde\Omega(t)\right]
-\left[\hat\Gamma,\tilde\Omega(t)\right]_+
+\frac{1}{2}\left\{
\tilde{H}_{\rm S},\tilde\Omega(t)
\right\}_{\mbox{\boldmath$\cal B$}}
-\frac{1}{2}\left\{
\tilde\Omega(t), \tilde{H}_{\rm S}
\right\}_{\mbox{\boldmath$\cal B$}}
\;,
\label{eq:ddt-Omega_matrix_W}
\end{equation}
where $\tilde\Omega(t)$ is
the phase-space-dependent non-normalized density matrix 
of the system
and 
\be
\mbox{\boldmath$\cal B$}=
\left[\begin{array}{cc} {\bf 0} & {\bf 1} \\
-{\bf 1} & {\bf 0}\end{array}\right]
\label{eq:B}
\ee
is the symplectic matrix written in block form. Using it, one can write
the Poisson bracket in the form adopted in Eq. (\ref{eq:ddt-qcrho_matrix}):
\be
\left\{\tilde{H}_{\rm S},\tilde\Omega(t)\right\}_{\mbox{\boldmath$\cal B$}}
\equiv
\sum_{I,J=1}^{2N}
\frac{\partial\tilde{H}_{\rm S}}{\partial X_I}
{\cal B}_{IJ}
\frac{\partial\tilde\Omega(t)}{\partial X_J}
\;.
\label{eq:poissonB}
\ee
The dimension of phase space is $2N$.

The trace of the phase-space-dependent density matrix
$\tilde{\rm T}{\rm r}$ is defined as
\be
\tilde{\rm T}{\rm r}\left(\tilde\Omega(t)\right) =
{\rm Tr}^\prime\int dX \tilde\Omega(X,t) \;,
\ee
where ${\rm Tr}^\prime$ denotes a partial trace over the quantum
dynamical variables $\hat x$ and $\int dX$ is the integral in phase space.
The trace $\tilde{\rm T}{\rm r}\left(\tilde\Omega(t)\right)$
obeys the equation of motion
\be
\frac{d}{dt}
\tilde{\rm T}{\rm r}\left(\tilde\Omega(t)\right)
=
-2\tilde{\rm T}{\rm r}\left(\hat{\Gamma}\tilde\Omega(t)\right)
\;.
\label{eq:ddt-qcTraceOm}
\ee
The result given in Eq. (\ref{eq:ddt-qcTraceOm}) is proven 
in Appendix \ref{app:qce-nherm}.

Let us now introduce a phase-space-dependent normalized density matrix
$\tilde\rho(X,t)$:
\be
\tilde\rho(X,t)=\frac{\tilde\Omega(X,t)}
{\tilde{\rm T}{\rm r}\left(\tilde\Omega(X,t)\right)} \;.
\label{eq:def-qc-rho}
\ee
Quantum statistical averages of an arbitrary phase-space-dependent
operator $\tilde\chi$ can now be calculated as
\be
\langle\tilde\chi\rangle_{(t)}=
\tilde{\rm T}{\rm r}\left(\tilde\rho(t)\tilde\chi\right)
\;.
\label{eq:rho-ps-averages}
\ee
The phase-space-dependent normalized density matrix
obeys the non-linear equation of motion below
\begin{equation}
\frac{\partial}{\partial t}\tilde\rho(t)=
-i\left[\tilde{H}_{\rm S},\tilde\rho(t)\right]
-\left[\hat\Gamma,\tilde\rho(t)\right]_+
+ 2\hat\rho(t)\tilde{\rm T}{\rm r}\left(\hat{\Gamma}
\tilde{\rho}(t)\right)
+\frac{1}{2}\left\{\tilde{H}_{\rm S},\tilde\rho(t)\right\}_{\mbox{\boldmath$\cal B$}}
-\frac{1}{2}\left\{\tilde\rho(t),\tilde{H}_{\rm S}\right\}_{\mbox{\boldmath$\cal B$}}
\;.
\label{eq:ddt-qcrho_matrix}
\end{equation}

\section{Non-Hermitian quantum systems set in a constant-temperature
phase space}
\label{sec:nhscltb}

Equations (\ref{eq:ddt-Omega_matrix_W}) and (\ref{eq:ddt-qcrho_matrix})
describe the dynamics of non-Hermitian quantum systems
embedded in phase space \cite{as_15}.
When considering open quantum systems in more general settings,
it is common to consider the effects of thermal fluctuations.
Phase space coordinates undergo thermal fluctuations when 
their microscopic equilibrium state is described
by the canonical distribution function.
In this ensemble the thermodynamic temperature is constant.
Within the framework of the operator-valued Wigner formulation of quantum mechanics
\cite{zhang-balescu,balescu-zhang,quasi_lie},
temperature constraints are efficiently formulated
by means of quasi-Lie brackets \cite{quasi_lie,b3,b4}. 
In the classical case, deterministic thermostats, 
such as the Nos\'e-Hoover chain thermostat \cite{nhc},
are also implemented by means of quasi-Lie brackets \cite{b1,b2,aspvg}.
Below, we briefly explain how this is achieved.
In \emph{silico}  (\emph{i.e.}, on the computer),
temperature control is realized by augmenting the dimensions of phase space 
using just a few additional degrees of freedom. The new higher dimensional phase space
is known as extended phase space.
We indicate points in the extended phase space by means of the symbol $X^{\rm e}$.
In the case of the Nos\'e-Hoover chain thermostat, upon introducing
two fictitious coordinates $\Lambda_1,\Lambda_2$
with their associated momenta $\Pi_1,\Pi_2$, we have
the following point in the extended phase space:
\be
X^{\rm e}=(Q,\Lambda_1,\Lambda_2,P,\Pi_1,\Pi_2) \;.
\ee
To lighten the notation,
the phase space coordinates of the Nos\'e-Hoover chain thermostat are denoted
as ${\cal Y}=(\Lambda_1,\Lambda_2,\Pi_1,\Pi_2)$. For the reader, the symbol
$\cal Y$ denotes a calligraphic $Y$.  
A generic operator acting  upon the extended phase space is identifiable by the apex e. 
Considering a classical system with a potential term $V(Q)$ and
Hamiltonian $H_{\rm B}(X) = P^2/2 + V(Q)$,
the extended phase space Hamiltonian
(see \cite{b1,b2,aspvg} and references therein) includes the 
Nosé-Hoover chain energy \cite{nhc,b1} and is defined as
\ba
H^{\rm e}(X^{\rm e}) 
&=& H_{\rm B}(X) +\sum_{K=1}^2\left(\frac{\Pi_K^2}{2\mu_K}\right)
+gk_{\rm B}T\Lambda_1+k_{\rm B}T\Lambda_2
\nonumber\\
&=&
H_{\rm B}(X) + H_{nhc}({\cal Y})\;,
\label{eq:defHe}
\ea
where $\mu_K$ are fictitious inertial parameters,
$T$ is the termodynamic temperature, $k_{\rm B}$ is the Boltzmann
constant, $N$ is the number of physical coordinates $Q$
that are thermalized.
The equation of motion in extended phase space can be written as:
\be
\dot X_K^{\rm e}=
\left\{X_K^{\rm e},H^{\rm e}\right\}_{\mbox{\boldmath$\cal B$}^{\rm e}}
=
\sum_{I,J=1}^{2(N+2)}
\frac{\partial X_K^{\rm e}}{\partial X_I^{\rm e}}
{\cal B}_{IJ}^{\rm e} 
\frac{\partial H^{\rm e}}{\partial X_J^{\rm e}}
=\sum_{J=1}^{2(N+2)}
{\cal B}_{KJ}^{\rm e}
\frac{\partial H^{\rm e}}{\partial X_J^{\rm e}}
\;,
\label{eq:quasi-Hamiltonian-eq-X^e}
\ee 
where the first and second equality define a quasi-Lie bracket.
The last equality enlighten the matrix structure of the
equations of motion.
This last form also allows one to compactly define the compressibility
of phase space:
\be
\kappa^{\rm e}\equiv\sum_{I=1}^{2(N+2)}\frac{\partial\dot X_I^{\rm e}}
{\partial X_I^{\rm e}}
=
\sum_{I,J=1}^{2(N+2)}
\frac{\partial{\cal B}_{IJ}^{\rm e}}{\partial X_I^{\rm e}}
\frac{\partial H^{\rm e}}{\partial X_J^{\rm e}}
=-N\frac{\dot\Pi_1}{\mu_1}-\frac{\dot\Pi_2}{\mu_2}\;,
\label{eq:nhc-kappa}
\ee
The antisymmetric matrix $\mbox{\boldmath$\cal B$}^{\rm e}$, entering
Eqs. (\ref{eq:quasi-Hamiltonian-eq-X^e}) and (\ref{eq:nhc-kappa}),
defined as
\be
\mbox{\boldmath$\cal B$}^{\rm e} 
=\left[
\begin{array}{cccccc}
0 & 0 & 0 & 1 & 0 & 0 \\
0 & 0 & 0 & 0 & 1 & 0 \\
0 & 0 & 0 & 0 & 0 & 1 \\
-1 & 0 & 0 & 0 & -P & 0 \\
0 & -1 & 0 & P & 0 & -\Pi_1 \\
0 & 0 & -1 & 0 & \Pi_1 & 0 
\end{array}
\right]
\;,
\label{eq:defBe}
\ee
is an antisymmetric phase-space-dependent matrix generalizing the
symplectic matrix in Eq. (\ref{eq:B}).
Equation (\ref{eq:quasi-Hamiltonian-eq-X^e}) is called a quasi-Hamiltonian equation
because, while conserving $H^{\rm e}(X^{\rm e})$, it cannot be
derived from the Hamiltonian formalism alone:
in order to write the equation of motion (\ref{eq:quasi-Hamiltonian-eq-X^e}),
together with $H^{\rm e}(X^{\rm e})$, one also needs the antisymmetric
matrix $\mbox{\boldmath$\cal B$}^{\rm e}$ in Eq. (\ref{eq:defBe}).
The equation of motion for the distribution function $f(X^{\rm e},t)$ is
\be
\frac{\partial}{\partial t}f(t)=
-\left\{f(t),H^{\rm e}\right\}_{\mbox{\boldmath$\cal B$}^{\rm e}}
-\kappa^{\rm e} f(t) \;.
\label{eq:dot_fdistrib}
\ee

%%%%%%%%%%%%%%%%%%%%%%%%%%%%%%%%%%%%%%%%%%%%%%%%%%%%%%%%%%%%%%%%%%%%%%%%
Given the discussion above, our task becomes that
of generalizing Eq. (\ref{eq:dot_fdistrib}) to a non-Hermitian quantum system.
The phase-space-dependent Hamiltonian of the quantum system
must be defined on the extended phase space of
the Nos\'e-Hoover Chain thermostat:
\ba
\tilde{\cal H}^{\rm e} &=& \hat H + H_{\rm B}(X)+ \tilde H_I
+H_{\rm nhc}({\cal Y}) -i\hat\Gamma \nonumber\\
&=& \tilde H_{\rm S}^{\rm e} -i\hat\Gamma
\;.
\ea
A phase-space-dependent non-normalized density matrix
$\tilde{\Omega}^{\rm e}(t)$ can be introduced as well.
We postulate that $\tilde{\Omega}^{\rm e}(t)$ obeys
the equation of motion
\be
\frac{\partial}{\partial t}\tilde{\Omega}^{\rm e}(t)=
-i\left[\tilde{H}_{\rm S}^{\rm e},\tilde{\Omega}^{\rm e}(t)\right]
-\left[\hat\Gamma,\tilde{\Omega}^{\rm e}(t)\right]_+
+\frac{1}{2}\left\{
\tilde{H}_{\rm S}^{\rm e},\tilde\Omega^{\rm e}
\right\}_{\mbox{\boldmath$\cal B$}^{\rm e}}
-\frac{1}{2}\left\{
\tilde\Omega^{\rm e},\tilde{H}_{\rm S}^{\rm e}
\right\}_{\mbox{\boldmath$\cal B$}^{\rm e}}
-\kappa_{\rm S}^{\rm e}\tilde{\Omega}^{\rm e}(t)
\;.
\label{eq:ddt-Omega_e_matrix}
\end{equation}
In absence of quantum degrees of freedom, Eq. (\ref{eq:ddt-Omega_e_matrix})
reduces correctly to the Eq. (\ref{eq:dot_fdistrib}).
Moreover, when $(\Lambda_I,\Pi_I)\to 0$, with $I=1,2$, (which means
that the Nos\'e-Hoover thermostat is not applied),
Eq. (\ref{eq:ddt-Omega_e_matrix}) reduces unerringly
to Eq. (\ref{eq:ddt-Omega_matrix_W}).

The trace $\tilde{\rm T}{\rm r}^{\rm e}$ of the density matrix 
$\tilde\Omega^{\rm e}(t)$ involves an integral over the 
extended phase space:
\be
\tilde{\rm T}{\rm r}^{\rm e}\left(\tilde{\Omega}^{\rm e}(t)\right) =
{\rm Tr}^\prime\int dX^{\rm e} \tilde\Omega^{\rm e}(X^{\rm e},t) \;.
\label{eq:treOmee}
\ee
The equation of motion of the trace defined in Eq. (\ref{eq:treOmee})
has the same structure of Eq. (\ref{eq:ddt-qcTraceOm}):
\be
\frac{d}{dt}
\tilde{\rm T}{\rm r}^{\rm e}
\left(\tilde{\Omega}^{\rm e}(t)\right)
=
-2\tilde{\rm T}{\rm r}^{\rm e}
\left(\hat{\Gamma}\tilde\Omega^{\rm e}(t)\right)
\;.
\label{eq:ddt-qcTraceOm_e}
\ee
In Appendix \ref{app:qce-nherm_nhc} it is shown 
how Eq. (\ref{eq:ddt-qcTraceOm_e}) is obtained.

The phase-space-dependent normalized density matrix in the extended phase space, $\tilde{\rho}^{\rm e}(X^{\rm e},t)$,
is defined in analogy with Eq. (\ref{eq:def-qc-rho}) as
\be
\tilde{\rho}^{\rm e}(X^{\rm e},t)=\frac{\tilde{\Omega}^{\rm e}(X^{\rm e},t)}
{\tilde{\rm T}{\rm r}^{\rm e}\left(\tilde{\Omega}^{\rm e}(X^{\rm e},t) \right)}
\;.
\label{eq:def-qc-rhoe}
\ee
In analogy with Eq. (\ref{eq:rho-ps-averages}), quantum statistical
averages are calculated as
\be
\langle\tilde\chi^{\rm e}\rangle_{(t)}^{\rm e}
=\tilde{\rm T}{\rm r}^{\rm e}\left(\tilde\rho^{\rm e}(t)
\tilde\chi^{\rm e}\right)\;.
\label{eq:rho-pse-averages}
\ee
The phase-space-dependent normalized density matrix 
in Eq. (\ref{eq:def-qc-rhoe}) obeys the non-linear equation of motion:
\ba
\frac{\partial}{\partial t}\tilde{\rho}^{\rm e}(t)
&=&
-i\left[\tilde H_{\rm S}^{\rm e},\tilde\rho^{\rm e}(t)\right]
-\left[\hat\Gamma ,\tilde\rho^{\rm e}(t)\right]_+
+2\tilde\rho^{\rm e}(t)
\tilde{\rm T}{\rm r}^{\rm e}\left(\hat\Gamma\tilde\rho^{\rm e}(t)\right)
\nonumber\\
&+&\frac{1}{2}\left\{
\tilde{H}_{\rm S}^{\rm e},\tilde\rho^{\rm e}
\right\}_{\mbox{\boldmath$\cal B$}^{\rm e}}
-\frac{1}{2}\left\{
\tilde\rho^{\rm e},\tilde{H}_{\rm S}^{\rm e}
\right\}_{\mbox{\boldmath$\cal B$}^{\rm e}}
-\kappa_{\rm S}^{\rm e}\tilde\rho^{\rm e}(t)
\;.
\label{eq:ddt-qcrho_matrix_e}
\ea
We remark that Eqs. (\ref{eq:ddt-Omega_e_matrix}) and (\ref{eq:ddt-qcrho_matrix_e})
are two important results of this work since these equations define the dynamics of a non-Hermitian quantum system embedded in a classical bath at constant temperature.

%%%%%%%%%%%%%%%%%%%%%%%%%%%%%%%%%%%%%%%%%%%%%%%%%%%%%%%%%%%%%%%%%%%%

\section{Model of a Non-Hermitian Quantum Single-Molecule Junction
at Constant Temperature}
\label{sec:dmjtb}

In this section we formulate a molecular junction toy-model described by a non-Hermitian Hamiltonian denoted by $\tilde{\cal H}^{\rm m}$. 
The various energy contributions entering the Hermitian terms in part of the Hamiltonian model are:
\ba
\hat H^{\rm m}&=& -\Delta\hat\sigma_z \;, \label{eq:HS}
\\
H_{\rm B}^{\rm m}(X) &=& \frac{P^2}{2}+\frac{\omega^2}{2}Q^2 \;,
\label{eq:HBm} 
\\
\tilde H_{\rm I}^{\rm m}(Q) &=& - c Q \hat \sigma_x \label{eq:HSB} \;, 
\label{eq:HIm}
\\
H_{\rm nhc}^{\rm m}({\cal Y})
&=&\sum_{K=1}^2\left(\frac{\Pi_K^2}{2\mu_K} + k_B T\Lambda_K\right) \;.
\label{eq:defHnhcm}
\ea
The two-level system Hamiltonian $\hat H^{\rm m}$ describes quantum transport
between the leads in terms of the variation of the population of a ground and an excited state. The extended phase space point $X^{\rm m}$ is
defined as in Eq. (\ref{eq:quasi-Hamiltonian-eq-X^e}) but the model
considers a single $Q$ coordinate and its conjugate momentum $\Pi$.
Transport in the isolated two-level system takes place through tunneling. The
inclusion in the model description of the decay operator $\hat\Gamma^{\rm m}$
opens a channel through which the population of the states can disappear
forever from the two-level system.
The explicit form of $\hat\Gamma^{\rm m}$ will be given later on when
we transform to the adiabatic basis.
This subsystem is also coupled to a harmonic mode, with free Hamiltonian
$H_{\rm B}^{\rm m}(Q)$, by means of an interaction Hamiltonian 
$\tilde H_{\rm I}^{\rm m}$.
The harmonic mode opens another channel for the transport of the
population between the leads. In addition to this, the harmonic
mode is embedded in a thermal bath \emph{via} the Nos\'e-Hoover chain thermostat
with energy $H_{\rm nhc}^{\rm m}$.
The complete non-Hermitian Hamiltonian model $\tilde{\cal H}^{\rm m}$ reads
\ba
\tilde{\cal H}^{\rm m}(X^{\rm m})
&=&
\hat H^{\rm m}+H_{\rm B}^{\rm m}(X)+\tilde H_{\rm I}^{\rm m}(Q)+
H_{\rm nhc}^{\rm m}({\cal Y})-i\hat\Gamma^{\rm m} 
\nonumber\\
&=&\tilde H_{\rm S}^{\rm m}(X^{\rm m})-i\hat\Gamma^{\rm m} 
\;, \label{eq:Hupsilon}
\ea
where the last equality also defines the Hermitian part of the Hamiltonian
model, \emph{i.e.}, 
\ba
\tilde H_{\rm S}^{\rm m}(X^{\rm m})
=\hat H^{\rm m}+H_{\rm B}^{\rm m}(X)+\tilde H_{\rm I}(Q)+
H_{\rm nhc}^{\rm m}({\cal Y})
\;.
\ea 
To the authors' knowledge, 
the Hamiltonian model in Eq. (\ref{eq:Hupsilon})
provides a novel non-linear approach to the modeling of
lossy molecular junctions in thermal baths.
Figure \ref{Fig1:ModelScheme} displays a pictorial representation of the 
non-Hermitian quantum single-molecule junction
(nHQSMJ) model at constant temperature.

The abstract dynamics of the nHQSMJ model is obtained upon
using the non-Hermitian Hamiltonian (\ref{eq:Hupsilon})
in Eq. (\ref{eq:ddt-qcrho_matrix_e}). The abstract equations of motion
can be represented using different basis sets. The approach described
in Ref. \cite{theochem} is based on the
representation in the adiabatic basis. We also adopt such an
approach here. The adiabatic Hamiltonian is defined as
\be
\tilde h_{\rm ad}^{\rm m}(Q) =
\hat H^{\rm m}+H_{\rm B}^{\rm m}(X)+
\tilde H_{\rm I}^{\rm m}(Q)-\frac{P^2}{2}
\;,
\ee
and the adiabatic basis is introduced through the eigenvalue equation
\be
\tilde h_{\rm ad}^{\rm m}(Q)|\Phi_\alpha(Q)\rangle
= E_\alpha |\Phi_\alpha(Q)\rangle \;.
\ee
In practice, the adiabatic states are a kind of \emph{dressed} states
of the quantum system when it interacts with approximately ``frozen''
classical degrees of freedom.
The operator $\hat\Gamma^{\rm m}$ is chosen in a phenomenological way.
In the adiabatic basis it is taken as
\be
\hat \Gamma^{\rm m}= \frac{\gamma}{2} \left[\begin{array}{cc} 1 & 0 \\ 0 & 0 \end{array}\right] \;.
\label{eq:constGamma}
\ee
In this way  we describe a model where the excited state is in contact with a sink while the ground state is left unperturbed.

The abstract equation of motion for the
non-normalized density matrix of the model is
\ba
\frac{\partial}{\partial t}\tilde\Omega^{\rm m}(t)
&=&
-i\left[\tilde H_{\rm S}^{\rm e},\tilde\Omega^{\rm m}(t)\right]
-\left[\Gamma^{\rm m} ,\tilde\Omega^{\rm m}(t)\right]_+
+\frac{1}{2}\left\{
\tilde{H}_{\rm S}^{\rm m},\tilde\Omega^{\rm m}
\right\}_{\mbox{\boldmath$\cal B$}^{\rm m}}
\nonumber\\
&-&\frac{1}{2}\left\{
\tilde\Omega^{\rm m},\tilde{H}_{\rm S}^{\rm m}
\right\}_{\mbox{\boldmath$\cal B$}^{\rm m}}
-\kappa_{\rm S}^{\rm m}\tilde\Omega^{\rm m}(t)
\nonumber\\
%%%%%%%%%%%%%%%%%%%%%%%%%%%%%%%%%%%
&=&
\left(-i\tilde{\cal L}^{\rm m}
-i\hat{\cal L}^{\Gamma^{\rm m}}\right)\tilde\Omega^{\rm m}(t)
\;.
\label{eq:liouville_op_OME}
\ea
The antisymmetric matrix $\mbox{\boldmath$\cal B$}^{\rm m}$ has the same structure
of that in Eq. (\ref{eq:defBe}), but it considers only the physical degrees
of freedom of the harmonic modes. Accordingly, the phase space compressibility
of the  model becomes $\kappa_{\rm S}^{\rm m}=-(\dot\Pi_1/\mu_1)-(\dot\Pi_2/\mu_2)$.

Equation (\ref{eq:liouville_op_OME}) introduces the two Liouville operators:
\ba
-i\tilde{\cal L}^{\rm m}
&=&
-i\left[\tilde H_{\rm S}^{\rm e},...\right]
+\frac{1}{2}\left\{
\tilde{H}_{\rm S}^{\rm m},...\right\}_{\mbox{\boldmath$\cal B$}^{\rm m}}
-\frac{1}{2}\left\{...,\tilde{H}_{\rm S}^{\rm m}
\right\}_{\mbox{\boldmath$\cal B$}^{\rm m}}
-\kappa^{\rm m}\left(...\right)
\;, \label{eq:liouv_op_Hm}
\\
-i\hat{\cal L}^{\Gamma^{\rm m}}
&=&
-\left[\Gamma^{\rm m},...\right]_+
\;.\label{eq:liouv_op_Gamma}
\ea
The equation of motion (\ref{eq:liouville_op_OME}) is represented into the
adiabatic basis as
\ba
\frac{d}{dt}\tilde{\Omega}_{\alpha\alpha'}^{\rm m}(t)
&=&
\sum_{\beta\beta'}
\left(-i\tilde{\cal L}_{\alpha\alpha',\beta\beta'}^{\rm m}
-i\hat{\cal L}_{\alpha\alpha',\beta\beta'}^{\Gamma^{\rm m}}\right)
\tilde\Omega_{\beta\beta'}^{\rm m}(t)
\;.
\label{eq:liouville_op_ad_bas}
\ea
In Eq. (\ref{eq:liouville_op_ad_bas}) we introduced
the two Liouville super-operators 
$-i\tilde{\cal L}_{\alpha\alpha',\beta\beta'}^{\rm m}$ and
$-i\hat{\cal L}_{\alpha\alpha',\beta\beta'}^{\Gamma^{\rm m}}$.
We first consider $\tilde{\cal L}_{\alpha\alpha',\beta\beta'}^{\rm m}$.
We have
\ba
-i\tilde{\cal L}^{\rm m}_{\alpha\alpha',\beta\beta'}
&=&
\left(-i\omega_{\alpha\alpha'} -  i \tilde L_{\alpha\alpha'}^{\rm m}
-\kappa^{\rm m} \right)
\delta_{\alpha\beta}\delta_{\alpha'\beta'}
+\tilde{\cal T}_{\alpha\alpha',\beta\beta'}\;,
\label{eq:iLaapbbpm}
\ea
where $\omega_{\alpha\alpha'}=E_\alpha(Q)-E_{\alpha'}(Q)$
is the Bohr frequency. The operator
\ba
i \tilde L_{\alpha\alpha'}^{\rm m}
&=&
P\frac{\partial}{\partial Q}
+ 
\frac{1}{2}\left(F_{\alpha}(Q) + F_{\alpha'}(Q) \right) 
\frac{\partial}{\partial P}
+
P\frac{\Pi_1}{\mu_1}\frac{\partial}{\partial P}
+ \nonumber\\
& &
- \frac{\Pi_1}{\mu_1} \frac{\partial}{\partial \Lambda_1}
- \left(P^2-k_BT\right)\frac{\partial}{\partial\Pi_1}
+\Pi_1\frac{\Pi_2}{\mu_2}\frac{\partial}{\partial\Pi_1}
+ \nonumber\\
& &
-\frac{\Pi_2}{\mu_2}\frac{\partial}{\partial\Lambda_2}
-\left(\frac{\Pi_1^2}{\mu_1}-k_BT\right)
\frac{\partial}{\partial\Pi_2}
\label{eq:nhc_Liouv_qc_op}
\ea
is a classical-like Liouville operator generating the dynamics
of the physical coordinates $X$ under
the feedback of the fictitious coordinates $\cal Y$.
In Eq. (\ref{eq:nhc_Liouv_qc_op})
we have also introduced the Hellmann-Feynman force:
\be
F_\alpha(Q)= - \frac{\partial E_\alpha(Q)}{\partial Q} \;.
\ee
This is the force acting on the harmonic oscillator when
state $\alpha$ is occupied.

The adiabatic surfaces transition operator $\tilde{\cal T}_{\alpha\alpha',\beta\beta'}$ 
is purely off-diagonal. In the adiabatic basis it is
expressed as
\ba
{\cal T}_{\alpha\alpha',\beta\beta'} 
&=&
\delta_{\alpha'\beta'} P \cdot {\cal C}_{\alpha\beta}(Q)
\left(1+\frac{1}{2}\tilde{\cal S}_{\alpha\beta}
\cdot\frac{\partial}{\partial P}\right)
\nonumber\\
&+&
\delta_{\alpha\beta}P\cdot {\cal C}_{\alpha'\beta'}^*(Q)
\left(1+\frac{1}{2}\tilde{\cal S}_{\alpha'\beta'}^*\cdot
\frac{\partial}{\partial P}\right)
\;,\nonumber
\ea
where
\ba
{\cal C}_{\alpha\beta}
&=&
\langle\Phi_\alpha(Q)|\frac{\partial}{\partial Q} |\Phi_\beta(Q)\rangle
\label{eq:Caap}
\;,
\\
\tilde{\cal S}_{\alpha\beta}
&=&
\frac{\left(E_{\alpha}-E_{\beta}\right)}{P\cdot {\cal C}_{\alpha\beta}(Q)}
{\cal C}_{\alpha\beta}(Q) \;.
\label{eq:Saap}
\ea
Equation (\ref{eq:Caap}) defines the adiabatic coupling vector 
${\cal C}_{\alpha\beta}$. This vector quantifies the superposition
of the adiabatic states when $Q$ changes.
Equation (\ref{eq:Saap}) introduces the vector $\tilde{\cal S}_{\alpha\beta}$.
If dimensional coordinates were introduced, $\tilde{\cal S}_{\alpha\beta}$
would have the dimension of momentum. Roughly speaking, 
$\tilde{\cal S}_{\alpha\beta}$ provides the momentum variation when jumping
from one adiabatic surface to the other.

We now consider $\hat{\cal L}^{\Gamma^{\rm m}}$. In general, a decay operator 
$\tilde\Gamma_{\alpha,\beta}$ can always be decomposed in a diagonal part
$
\tilde\Gamma_{\alpha\beta}^{\rm d}
=\tilde\Gamma_{\alpha\beta}\delta_{\alpha\beta}
\;,
$
and an off-diagonal part
$
\tilde\Gamma_{\alpha\beta}^{\rm o}=\tilde\Gamma_{\alpha\beta}
\left(1-\delta_{\alpha\beta}\right)\;,
$
so that
$
\tilde\Gamma_{\alpha,\beta}\equiv \tilde\Gamma_{\alpha\beta}^{\rm d}
+ \tilde\Gamma_{\alpha\beta}^{\rm o} \;.
$
However, in the case of the decay operator of the nHQSMJ model adopted here,
and defined in Eq. (\ref{eq:constGamma}), we have 
$\left(\tilde\Gamma^{\rm m}\right)_{\alpha\beta}^{\rm o} = 0$.
Hence, we have
\be
i{\cal L}_{\alpha\alpha',\beta\beta'}^{\Gamma^{\rm m}}
=\gamma_{\alpha\alpha'}^{\rm m}\delta_{\alpha\beta}\delta_{\alpha'\beta'}
\label{eq:iLaapbbpgm}
\;,
\ee
where $\gamma_{\alpha\alpha'}^{\rm m}=\Gamma_{\alpha\alpha}^{\rm m}
+\Gamma_{\alpha'\alpha'}^{\rm m}$.

Using Eqs. (\ref{eq:iLaapbbpm}) and (\ref{eq:iLaapbbpgm}),
Eq. (\ref{eq:liouville_op_ad_bas}) becomes
\ba
\frac{d}{dt}\tilde{\Omega}_{\alpha\alpha'}^{\rm m}(t)
&=&
-\sum_{\beta\beta'}
\left[\left(
i\omega_{\alpha\alpha'} + \gamma_{\alpha\alpha'}^{\rm m}
+i\tilde L_{\alpha\alpha'}^{\rm m} +\kappa_{\rm S}^{\rm m }\right)\delta_{\alpha\beta}\delta_{\alpha'\beta'}
+\tilde{\cal T}_{\alpha\alpha',\beta\beta'} \right]
\tilde\Omega_{\beta\beta'}^{\rm m}(t)
\;.
\label{eq:liouville_opm_ad_bas}
\ea
Equation (\ref{eq:liouville_opm_ad_bas})
provides the equation of motion for the density matrix of the nHQSMJ model
at constant temperature.
This equation can be implemented through a variety of algorithms
\cite{theochem,sstp,algo12,algo15}.

\section{Calculations and Results}
\label{sec:results}

We consider a situation where the molecule is weakly coupled to
the leads. This implies that neglecting
the action of the transition operator
 $\tilde{\cal T}_{\alpha\alpha',\beta\beta'}$
in Eq. (\ref{eq:liouville_opm_ad_bas})
the phase-space-dependent density matrix is propagated 
by means of the sequential short-time propagation (SSTP)
algorithm \cite{theochem,sstp} as 
\ba
\tilde{\Omega}_{\alpha\alpha'}^{\rm m}(t)
&=&
\prod_{j=1}^{n_{\rm step}}
\sum_{\alpha\alpha'}
e^{ -i\left(\omega_{\alpha\alpha'} -i \gamma_{\alpha\alpha'}^{\rm m}
-i\kappa^{\rm m}
+\tilde L_{\alpha\alpha'}^{\rm m}\right) \tau }
\tilde{\Omega}_{\alpha\alpha'}^{\rm m},(0)
\nonumber\\
&=&
\prod_{j=1}^{n_{\rm step}}
\sum_{\alpha\alpha'}
e^{ -i \tilde{\cal L}_{\alpha\alpha'}^{\rm m,(0)} \tau }
\tilde{\Omega}_{\alpha\alpha'}^{\rm m}(0)
\;,
\ea
where $\tau=t/n_{\rm step}$ and the non-Hermitian adiabatic
Liouville super-operator is
$ -i \tilde{\cal L}_{\alpha\alpha'}^{\rm m(0)} =
-i\omega_{\alpha\alpha'}-\gamma_{\alpha\alpha'}^{\rm m}
-\kappa_{\rm S}^{\rm m} - i \tilde L_{\alpha\alpha'}^{\rm m}$.
Quantum statistical averages can be calculated as
\ba
\tilde{\rm T}{\rm r}^{\rm m}\left(\tilde\rho^{\rm m}(t)\tilde\chi^{\rm m}\right)
&=&
\frac{
\sum_{\alpha\alpha'}\int dX^{\rm m}
\tilde{\Omega}_{\alpha\alpha'}^{\rm m}(X^{\rm m},t)
\tilde \chi_{\alpha'\alpha}(X^{\rm m})
}{ \sum_{\sigma}\int dX^{\rm m}
\tilde{\Omega}_{\sigma\sigma}^{\rm m}(X^{\rm m},t) }
\nonumber\\
&=&
\frac{
\sum_{\alpha\alpha'}\langle \tilde{\Omega}_{\alpha\alpha'}^{\rm m}(t)
\tilde \chi_{\alpha'\alpha} \rangle^{\rm m}
}{ \sum_{\sigma}\langle\tilde{\Omega}_{\sigma\sigma}^{\rm m}(t)
\rangle^{\rm m} } \;, \label{eq:stsa_ave}
\ea
where the bracket $\langle ... \rangle^{\rm m}$ stands for the average
in the extended phase space of the model.
We implement Eq. (\ref{eq:stsa_ave}) in the following way \cite{theochem,sstp}. 
Once the quantum initial state is assigned for every
point of the extended phase space, the calculation of quantum averages
can be performed by sampling the initial $X^{\rm e}$ with a 
Monte Carlo algorithm.
The point $X^{\rm m}$ is then propagated
over the assigned quantum state for a time length $t$.
The time step $\tau=t/n_{\rm step}$ in Eq. (\ref{eq:stsa_ave})
must be chosen small enough
to minimize the numerical error. 
In the calculation we take the following uncorrelated form
of the initial phase-space-dependent density matrix
\be
\tilde\Omega^{\rm e}(X^{\rm m},0)
=\hat\Omega^{\rm m}(0)\Omega_{\rm B}^{\rm m}(X,0)
\Omega_{\rm nhc}^{\rm m}({\cal Y},0)\;,
\ee
where
\ba
\hat\Omega^{\rm m}(0)
&=&\left[\begin{array}{cc} 0 & 0 \\ 0 & 1 \end{array}\right] \;,
\\
\nonumber\\
\Omega_{\rm B}^{\rm m}(X,0)
&=&\frac{\tanh(\beta\omega/2)}{\pi}\exp
\left(-\frac{\tanh(\beta\omega/2)}{\omega} H_{\rm B}^{\rm m}(X)\right)\;,
\\
\Omega_{\rm nhc}^{\rm m}({\cal Y},0)
&=&
\frac{\prod_{J=1}^2\delta(\Lambda_J)\exp(-\Pi_J^2/2)}
{Z_{\cal Y}^{\rm m}}
\ea
and
\be
\int d{\cal Y}\prod_{J=1}^2\delta(\Lambda_J)\exp(-\Pi_J^2/2\sigma^2)
\ee

Calculations were performed with fixed values of $\gamma=0.1$, $\omega=1/3$,
$\Delta=1$, $c=0.007$, time step $\tau=0.005$, number of time steps 
$n_{\rm step} = 10^4$, and number of Monte Carlo steps $n_{\rm mcs} = 25 \cdot 10^3$.
All the values of the parameters are given in adimensional units.
Twenty different values of the inverse temperature were considered upon chosing
 $\beta_{l+1}=1/T_{l+1}=\beta_0(1 + l ) $,
with $l=0,...,19$ and $\beta_0=0.0005$.
For each $\beta_l$ we have investigated four different types of dynamics:
i) Unitary dynamics. ii) Constant-temperature dynamics.
iii) Non-unitary dynamics. iv) Non-unitary dynamics at constant temperature.
Initial conditions are chosen with the two-level system
in its lowest eigenvector and the bath  in a thermal state.
 It is useful to introduce $\hat\Xi^m$ and 
$\hat{\cal X}^{\rm m}$, the non-normalized 
and normalized reduced density matrices, respectively.
We write their matrix elements in the adiabatic
basis as
\ba
\Xi_{\alpha\alpha'}^m &=& \int dX^{\rm m}\tilde\Omega_{\alpha\alpha'} \label{eq:Xi_aap^m}\;,
\\
{\cal X}_{\alpha\alpha'}^m &=& \int dX^{\rm m}\tilde\rho_{\alpha\alpha'}\label{eq:Chi_aap^m}\;.
\ea
Both reduced density matrices can be easily calculated
by means of our numerical approach.

In order to check the numerical scheme, we calculated the evolution in time
of $\tilde{\rm T}{\rm r}(\tilde\Omega^{\rm m}(t))$.
The results are shown in Fig. \ref{FIG2:Trace}.
When the decay operator is set to zero, Panels (a) and (b),
the trace of the density matrix is conserved with extremely good numerical
precision both when the dynamics is unitary
and when the temperature is controlled.
This provides a convincing indication that the
our algorithm conserves important dynamical invariants.
Panels (c) and (d) in Fig. \ref{FIG2:Trace} show the decrease in time
of $\tilde{\rm T}{\rm r}(\tilde\Omega^{\rm m}(t))$ 
non-unitary dynamics.Moreover, there is no appreciable difference between
unitary and constant-temperature dynamics.

In Fig. (\ref{FIG3:Rho}) we show the time-evolution of
$\Xi_{\alpha\alpha}^{\rm m}(t)$ and
${\cal X}_{\alpha\alpha}^{\rm m}(t)\rangle^{\rm m}$, with $\alpha=1,2$,
defined in Eq. (\ref{eq:Xi_aap^m}) and \mbox{(\ref{eq:Chi_aap^m})}, respectively.
Panels (a) and (b) display $\Xi_{11}^{\rm m}(t)$
and $\Xi_{22}^{\rm m}(t)$, respectively.
Initially, both adiabatic states are equally occupied.
However, according to Eq. (\ref{eq:constGamma}), only the occupation of
the excited state experiences depletion, see in Panel (a).
Panels (c) and (d) of \mbox{Fig. (\ref{FIG3:Rho})} show ${\cal X}_{11}^{\rm m}(t)$
and ${\cal X}_{22}^{\rm m}(t)$, respectively.
The behaviour of ${\cal X}_{11}^{\rm m}(t)$
is hardly distinguishable from that of $\Xi_{11}^{\rm m}(t)$.
However, the situation is different for ${\cal X}_{22}^{\rm m}(t)$ in Panel (d).
As a matter of fact, while $\Xi_{22}^{\rm m}(t)$
is constant, ${\cal X}_{22}^{\rm m}(t)$
increases so that the $\tilde{\rm T}{\rm r}^{\rm m}(\tilde\rho^{\rm m}(t))=
{\cal X}_{11}^{\rm m}(t)+
{\cal X}_{22}^{\rm m}(t)=1$.

In Fig. (\ref{FIG4:Rho12}) we show the real part of the reduced off-diagonal element
${\rm Re}({\cal X}_{12}^{\rm m}(t))$.
For the chosen values of the parameters, dephasing occurs on the same time scale of the
population's depletion of the excited state.

Figure (\ref{FIG5:SigmaTot_0.0075}) displays the time evolution of the
population difference for $\beta=0.0075$.
Below each plot showing the time evolution of the population difference,
the Fourier transform is also displayed.
The Fourier transform of Panel (a) displays two peaks:
one at small frequency and one at higher frequency.
The Rabi-like oscillations of the population difference at distant time
are signs of the absence of stable quantum transport in the nHQSMJ model.
Nos\'e-Hoover chain dynamics, shown in Panel (b), suppresses the peak
at small frequency.
The Rabi-like oscillations of the population difference at distant time
are less pronounced: they take place at relatively high frequency, as
the Fourier transform shows, around zero. This indicates that
in the presence of thermal noise the population difference
oscillates around zero so that transport is stable.
This can be seen as a particular instance of environment-assisted
quantum transport: remarkably, the noise provided by an environment
can defeat Anderson localization \cite{anderson} and assist quantum transport.
The idea that the noise of the environment could
enhance quantum transport, instead of suppressing it,
has been postulated to explain the high efficiency of energy transport
in  photosynthetic systems \cite{plenio}.
Obviously, the noise-enhancement of quantum transport can take place only
below a certain threshold. Above it, transport is suppressed by
the quantum Zeno effect \cite{zeno}.
In Panel (c) Hamiltonian dynamics of the bath is combined with 
a non-zero decay operator acting on the two-level system.
As it can be seen, the decay operator suppresses the high frequencies
in the Rabi-like oscillations of the population difference at distant time.
Such oscillations show that, even in this case, quantum transport
is not very stable. Comparison of the dynamics in Panel (b)
and in Panel (c) shows that there is a difference between thermal
and probability sink dissipation: the first suppresses low frequency
oscillation while the second damps high frequency oscillations.
As one can expect the combination of the two dynamics, shown
in Panel (d), leads to the suppression of both frequencies,
increasing the efficiency and the stability of the population transport.
Calculations for $\beta = 0.0025$ are reported
in Fig. (\ref{FIG7:SigmaTot_0.0025}). The results are indistinguishable
to the human eye from those obtained at $\beta=0.0075$.
This reinforces the conclusion that quantum transport in the nHQSMJ model
is enhanced by coupling the two-level system with a probability sink and 
controlling the temperature of the molecule.

\section{Conclusions}
\label{sec:conclusion}

In this work, we construct a theory for studying non-Hermitian
phase-space-dependent quantum systems at constant temperature.
This theory is based on an operator-valued Wigner formulation
of quantum mechanics or, in other words, on a phase-space-dependent density matrix.
The condition of constant temperature in phase space
is achieved by means of the Nos\'e-Hoover chain thermostat.
A mathematical result of the formalism is the derivation
of the non-linear equation of motion
for the normalized phase-space-dependent density matrix of the system.
We remark that this result, considering a constant temperature bath, improves the theory developed in Ref. \cite{as_15} where the temperature fluctuations are not constrained.
These latter situation is somewhat unrealistic in common experiments.

This theory is applied to a model of a non-Hermitian quantum single-molecule junction.
We emphasize that this model treats probability loss
and thermal fluctuations on the same level.
In detail, the model comprises
a two-level system with a probability sink coupled to
a thermalized harmonic mode.
The structure of our model was conjectured 
upon drawing an analogy with the process of noise-assisted transport
\cite{noise,noise2,noise3}.
In our case, we expect that the assisted quantum transport 
arises from the combined action of the probability sink and the
constant-temperature fluctuating molecule.
Indeed, we observe a transport enhancement \emph{in silico} upon
simulating numerically the non-Hermitian quantum single-molecule junction model
for a range of temperatures, while keeping
the values of the other parameters constant.
The Fourier transformed signal in frequency space shows
that the Nos\'e-Hoover chain thermostat suppresses the slow frequencies
while the probability sink damps the high frequency oscillations of
the population difference of the non-Hermitian quantum single-molecule junction 
model at constant temperature.

The non-Hermitian quantum single-molecule junction model at constant temperature
 introduced in this paper can 
more accurately be classified as a particular instance of a class of models.
As a matter of fact, it can be easily generalized
by considering a greater number of quantum states ($\sim 10$)
and an even greater number of classical modes ($\sim 10^3-10^5$).
Of course, more sophisticated algorithms
would have to be used for calculating the evolution of the density matrix.
We also note that spin models that are similar but simpler than the 
non-Hermitian quantum single-molecule junction model
introduced in this paper can be solved analytically
 when obeying certain symmetry properties
\cite{hn_16,gv_17,gm_17,gm_18,gm_19,gv_19,gv2_19,hn_18,hn2_18}.
It would be interesting to investigate whether the extension of
these models to non-Hermitian quantum mechanics would still be analytically treatable.
We defer such generalizations to future work.

%%%%%%%%%%%%%%%%%%%%%%%%%%%%%%%%%%%%%%%%%%%%%%%%% FIGURES %%%%%%%%%%%%%%%

%%%%%%%%%%%%%%%%%%%%% FIGURE 1 %%%%%%%%
\begin{figure}[H]
\includegraphics[width=14cm,height=12cm]{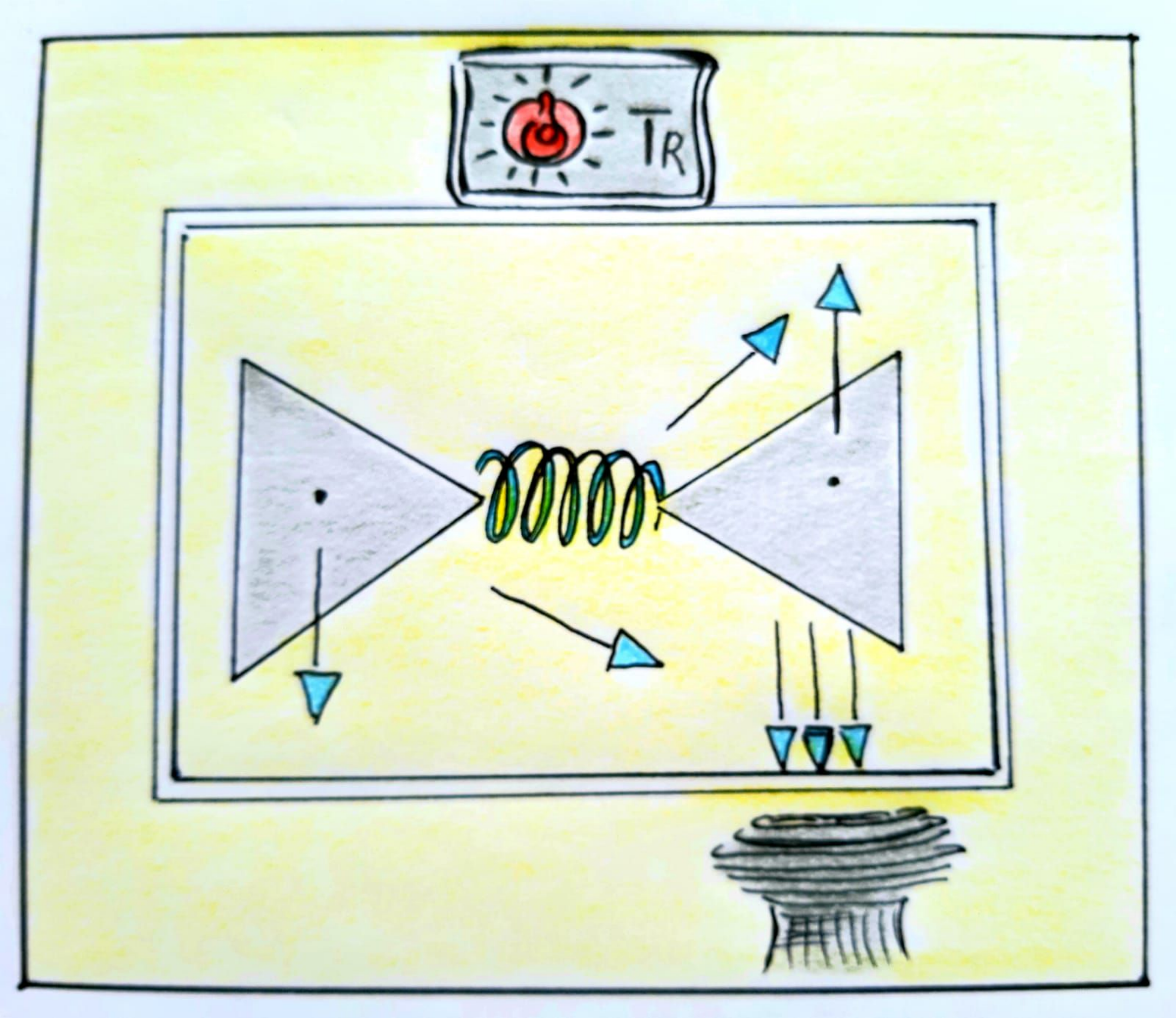}
\caption{Pictorial representation of the nHQSMJ model.
The leads are represented by two gray triangles.
The spring portrays the molecule connecting the leads.
The arrows with the cyan cusps depict the state of the TLS, depending on
its position along the junction. When the TLS is in the left lead, the arrow
points downward. As the TLS moves toward the right lead, the arrow rotates
until it reaches the upward position, which implies a complete transfer
to the right lead.
The sink below the right lead absorbs the TLS probability in an irreversible
way. In the formalism, the sink is represented by a Hermitian decay operator.
The decay operator acts on quantum states that in turn determine probabilities.
Hence, the action of the sink unfolds upon an ensemble of trajectories of the 
TLS system. The picture also shows a box separating the molecular junction
from the environment. The drawing of the box resembles on purpose
that of an oven with a thermostat on top. The thermostat temperature
is set at the same value of the $T_{\rm R}$ temperature of the environment.
The thermostat controls the temperature inside the oven so that it
is equal to that of the environment.
The yellow color inside the oven and around it, in the environment,
conveys the idea of thermal equilibrium.
} \label{Fig1:ModelScheme}
\end{figure}

%%%%%%%%%%%%%%%%%%%%% FIGURE 2 %%%%%%%%
\begin{figure}[H]
\includegraphics[width=14cm,height=12cm]{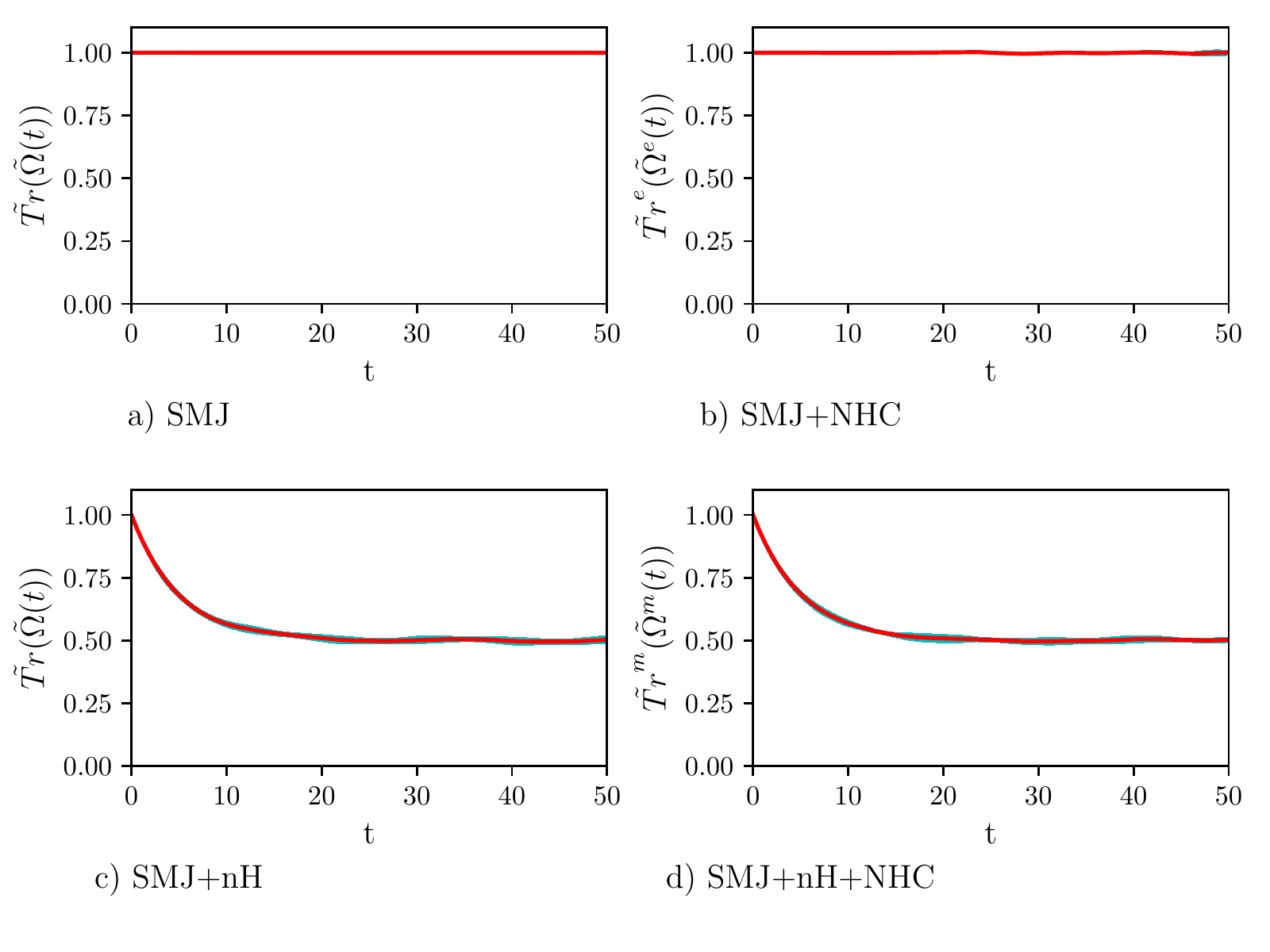}
\caption{Plot of $\tilde{\rm T}{\rm r}^{\rm}(\tilde\Omega^{\rm}(t))$.
Number of Monte Carlo sampled phase space initial conditions $X^{\rm e}$
points $=2500$, time step $\tau=0.005$, $\gamma = 0.1$, $\omega = 1/3$,
$\Delta = 1$, coupling constant $c = 0.007$, inverse temperature
$\beta = 0.0050$. The red line shows the average trend while the cyan area
around it, which is hardly visible, displays the statistical error of
the calculation. Panels (a) and (b) show the results of the calculations
when the dynamics takes place for the isolated single-molecule junction (SMJ)
and for the SMJ at constant temperature (SMJ+NHC), respectively.
Panel (c) shows the result of the calculation when a decay operator
acts onto the SMJ (SMJ+nH). 
Finally, in Panel (d) all types of dynamics are considered together
(SMJ+nH+NHC).
It is easy to see that the trace is conserved when the Hamiltonian is 
Hermitian. Otherwise, the population drops to about one half  as the 
probability is subtracted only from the excited state.} \label{FIG2:Trace}
\end{figure}

%%%%%%%%%%%%%%%%%%%%% FIGURE 3 %%%%%%%%
\begin{figure}[H]
\includegraphics[width=14cm,height=12cm]{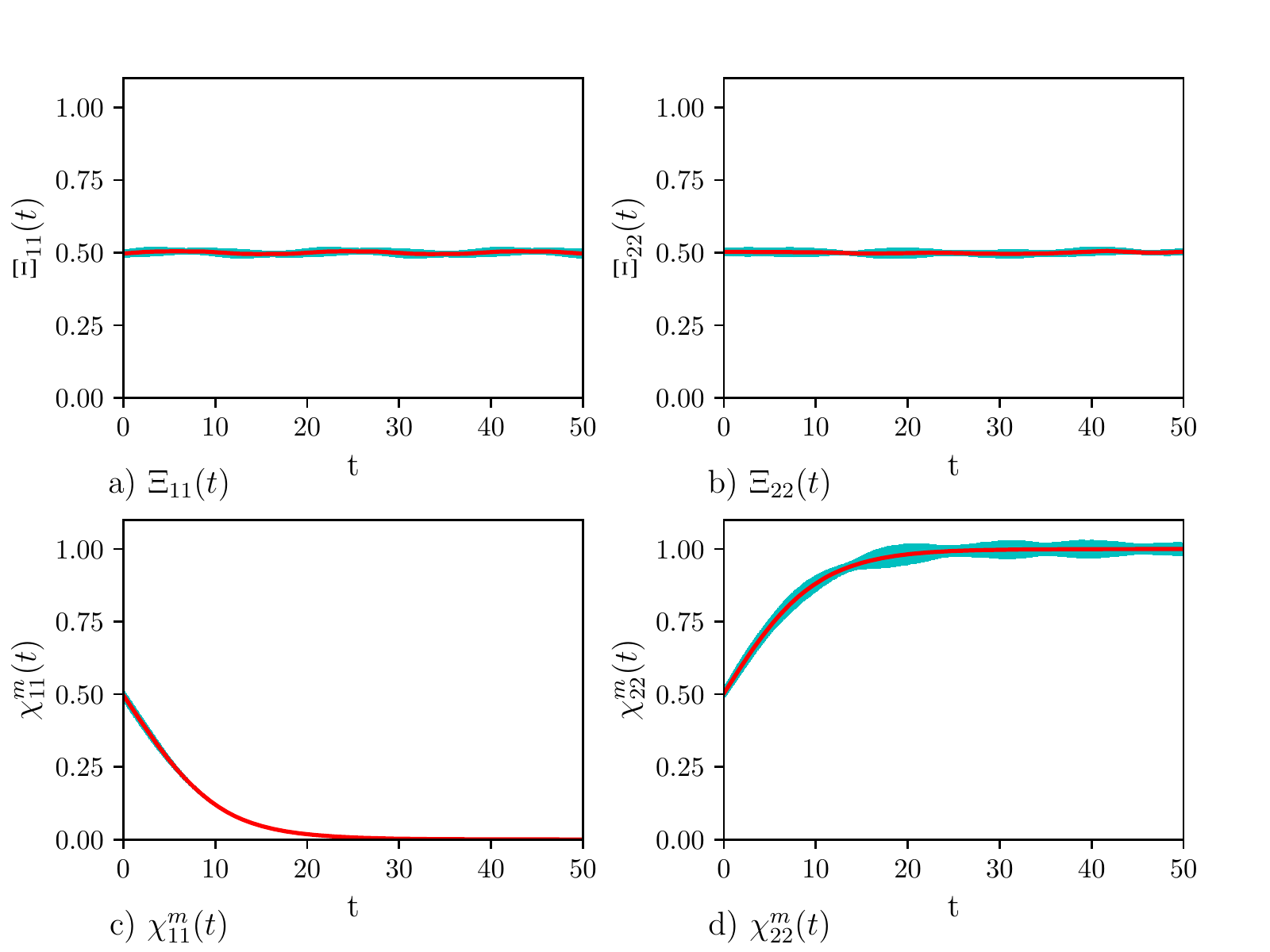}
\caption{The plots show the diagonal elements of the density matrices
resulting from nH+NHC dynamics.
Panels (a) and (b) show the behavior of the non-normalized density matrix elements $\Xi_{11}(t)$ and $\Xi_{22}(t)$, respectively.
While Panels (c) and (d) display the behavior of the normalized density matrix elements ${\cal X}_{11}^{\rm m}(t)$ and 
${\cal X}_{22}^{\rm m}(t)$ .
Number of iterations
$=10000$, number of Monte Carlo points $=2500$, time step $\tau=0.005$,
$\gamma = 0.1$, $\omega = 1/3$, $\Delta = 1$, $c = 0.007$, $\beta = 0.0050$.
The red line is the average trend, the cyan area (almost invisible to the eye)
 is the error. In (a) 
the effects of the decay operator $\hat{\Gamma}^{\rm m}$ can be observed. The trend in
(b) clarifies the cause for the trend of the trace in Fig.(\ref{FIG2:Trace}).
In (c) the difference with (a) is not evident, as the trace decreases directly
because of the decrease of $\Omega_{11}$. The comparison between (d) and (b)
further clarifies that the cause for the population loss of the excited state
is the decay operator and not energy transfer.} 
\label{FIG3:Rho}
\end{figure}

%%%%%%%%%%%%%%%%%%%%% FIGURE 4 %%%%%%%%
\begin{figure}[H]
\includegraphics[width=14cm,height=12cm]{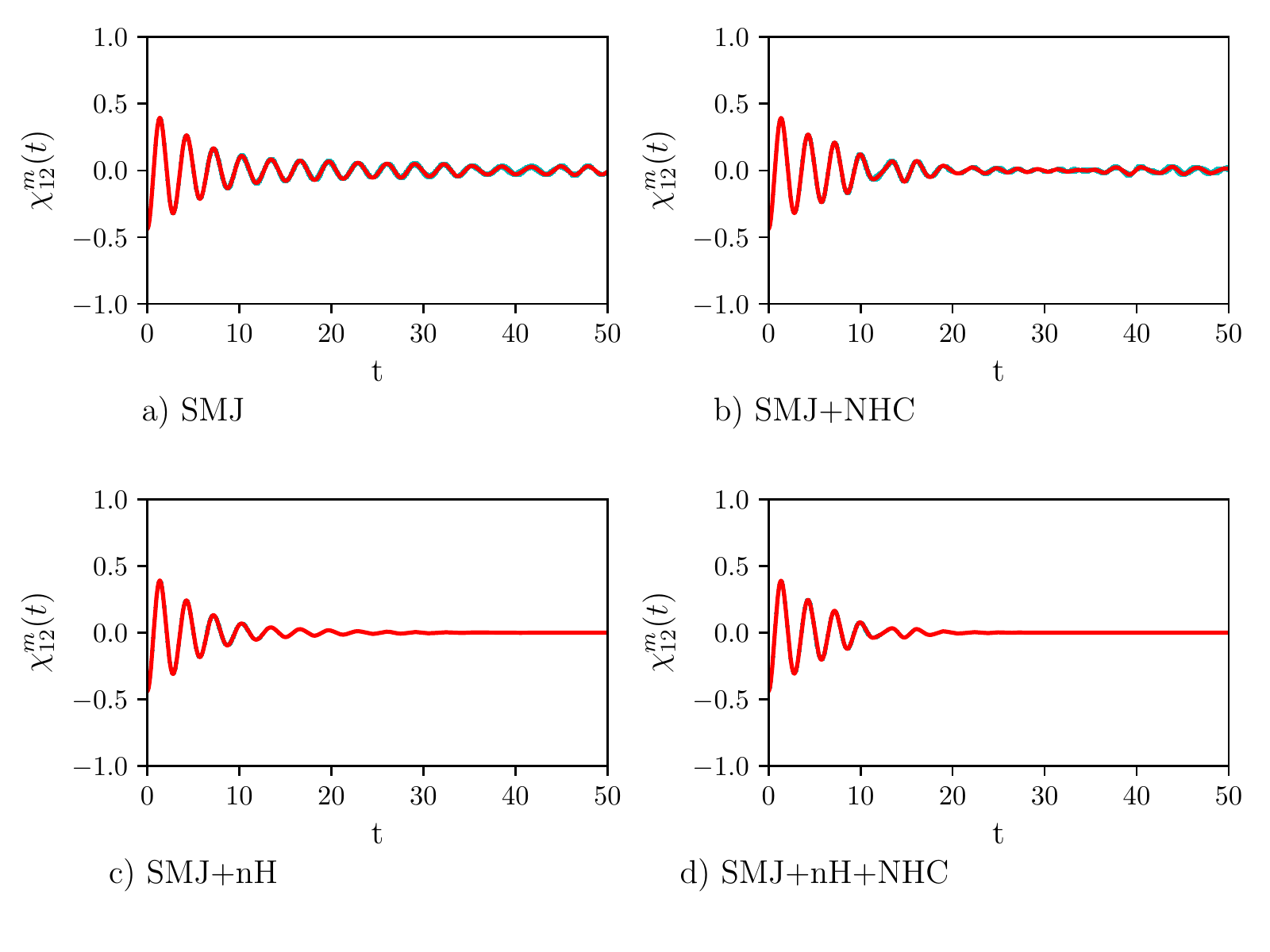}
\caption{The plot shows the average of the real part of 
${\cal X}_{12}^{\rm m}(t)$, the off-diagonal
element of the reduced normalized density matrix
(NHC+NH system). Number of iterations $=10000$, number of Monte Carlo points
$=2500$, time step $\tau=0.005$, $\gamma = 0.1$, $\omega = 1/3$, $\Delta = 1$,
$c = 0.007$, $\beta = 0.0050$. The red line is the average trend, the cyan area
is the error. The trend clearly shows decoherence.} 
\label{FIG4:Rho12}
\end{figure}

%%%%%%%%%%%%%% FIGURE5 %%%%%%%%%%%%%%%
\begin{figure}[H]
\includegraphics[width=14cm]{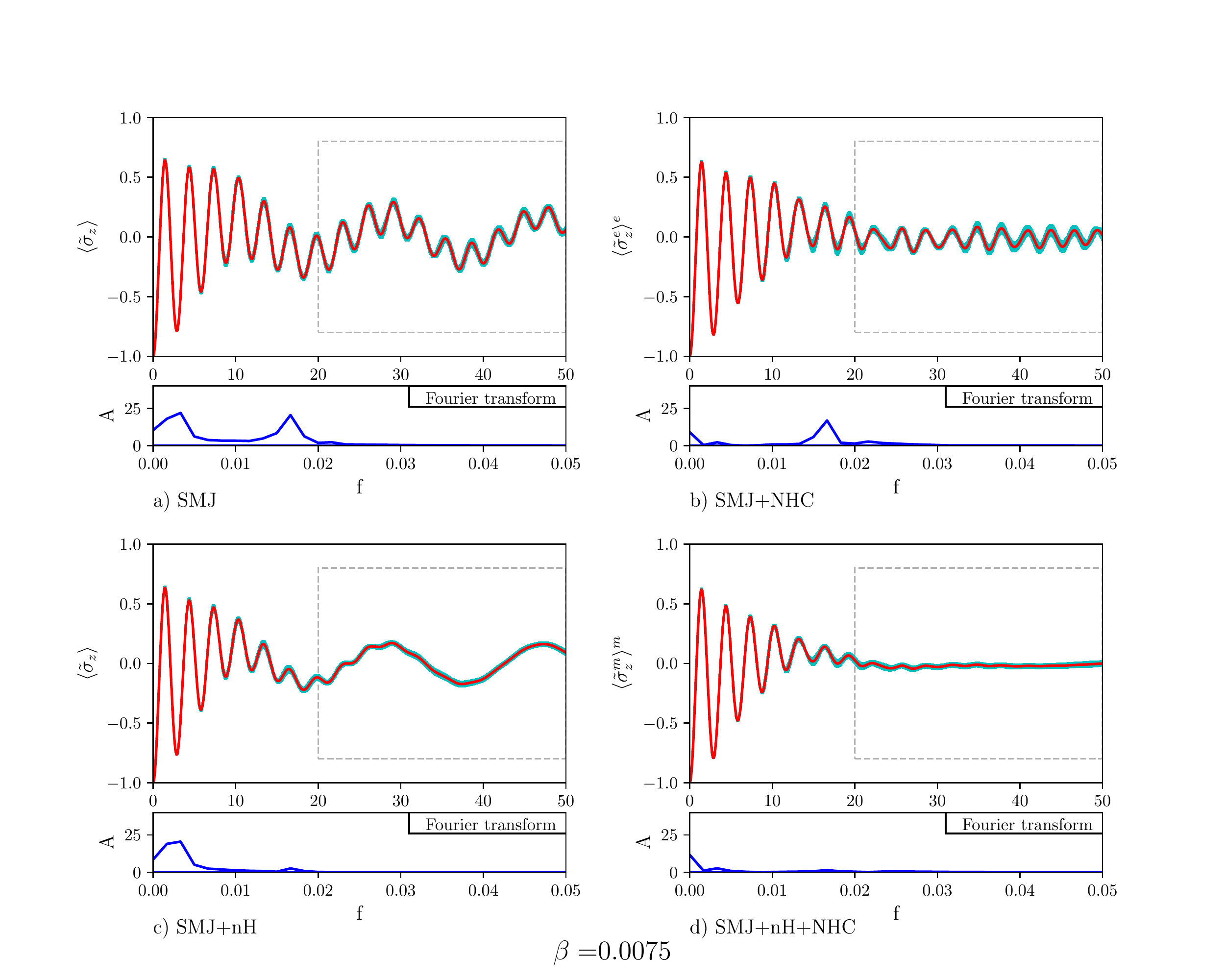}
\caption{
The plot shows the average population transfer
vs. time for different types of dynamics.
The values of the parameters of the calculation are
(NDS=10000, NMCP=2500, $\tau = 0.005$, 
$\gamma=0.1$, $\omega = 1/3$, $\Delta = 1$, $g = 0.007$,
$\beta = 0.0075$).
The red line indicates the average trend, the cyan area represents the error.
The blue line denotes the Fourier transform of the time evolution
of the average value (the data are considered after $t=20$).
Panel (a) displays the results for the isolated SMJ.
One can observe that the values oscillate with low and high frequency.
Panel (b) displays the dynamics of the SMJ with temperature control
(SMJ+NHC). The thermostat facilitates the population transfer
and stabilizes its average.
The low frequency terms are highly damped while the high frequency
ones are unaltered.
Panel (c) shows that the effect of
the decay operator (SMJ+nH) is to cancel the high frequency term.
Finally, in (d), it is shown that the union of the
NHC thermostat and decay operator (SMJ+nH+NHC)  determines a stable
tranfer.
Both the low and the high frequency terms are strongly damped.
} 
\label{FIG5:SigmaTot_0.0075}
\end{figure}

%%%%%%%%%%%%%%% FIGURE 6 %%%%%%%%%%%%%%
\begin{figure}[H]
\includegraphics[width=14cm]{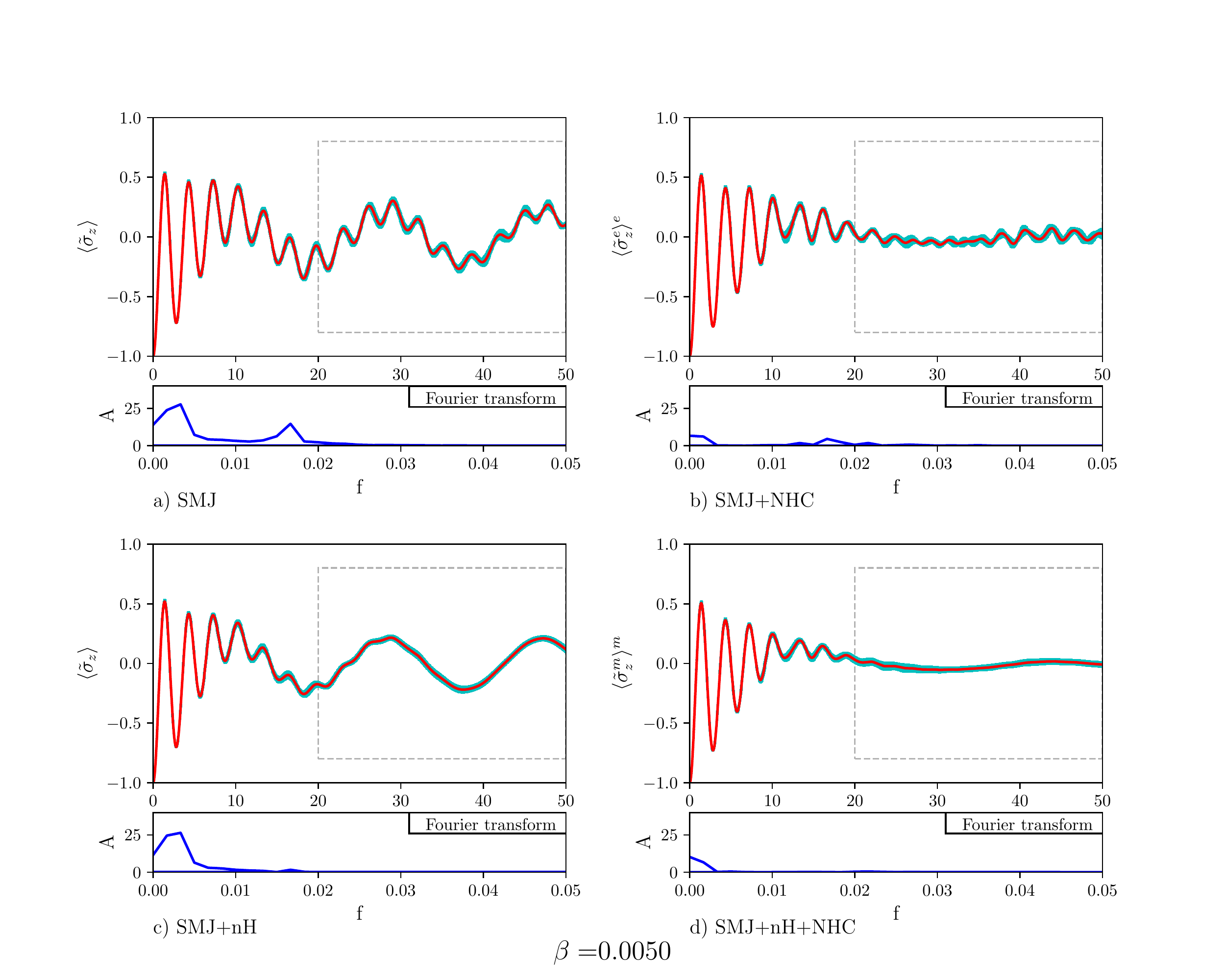}
\caption{The plot shows the average population transfer
vs. time for different types of dynamics.
The values of the parameters of the calculation are
(NMDS= 10000, NMCP = 2500, $\tau = 0.005$,  $\gamma = 0.1$,
$\omega$ = 1/3, $\Delta$ = 1, g = 0.007, $\beta$ = 0.0050).
The red line indicates the average trend, the cyan area represents the error.
The blue line denotes the Fourier transform of the time evolution
of the average value (the data are considered after $t=20$).
Panel (a) displays an average population transfer for the SMJ model where
low frequency oscillations are strongly damped.
In Panel (b) it is shown that a
higher temperature damps high frequency oscillations (SMJ+NHC).
Statistical error appears to be somewhat greater.
Panel (c) shows that the effect of the decay operator
is to damp high frequency oscillations (SMJ+nH).
Panel (d) shows that the combined action of the NHC thermostat 
and the decay operator (SMJ+nH+NMJ) determines a stable transfer process.
Both low and high frequency oscillations are strongly damped,
except for a very weak oscillation at a very
low frequency.
} 
\label{FIG6:SigmaTot_0.0050}
\end{figure}

%%%%%%%%%%%%% FIGURE 7 %%%%%%%%%%%%%%%
\begin{figure}[H]
\includegraphics[width=14cm]{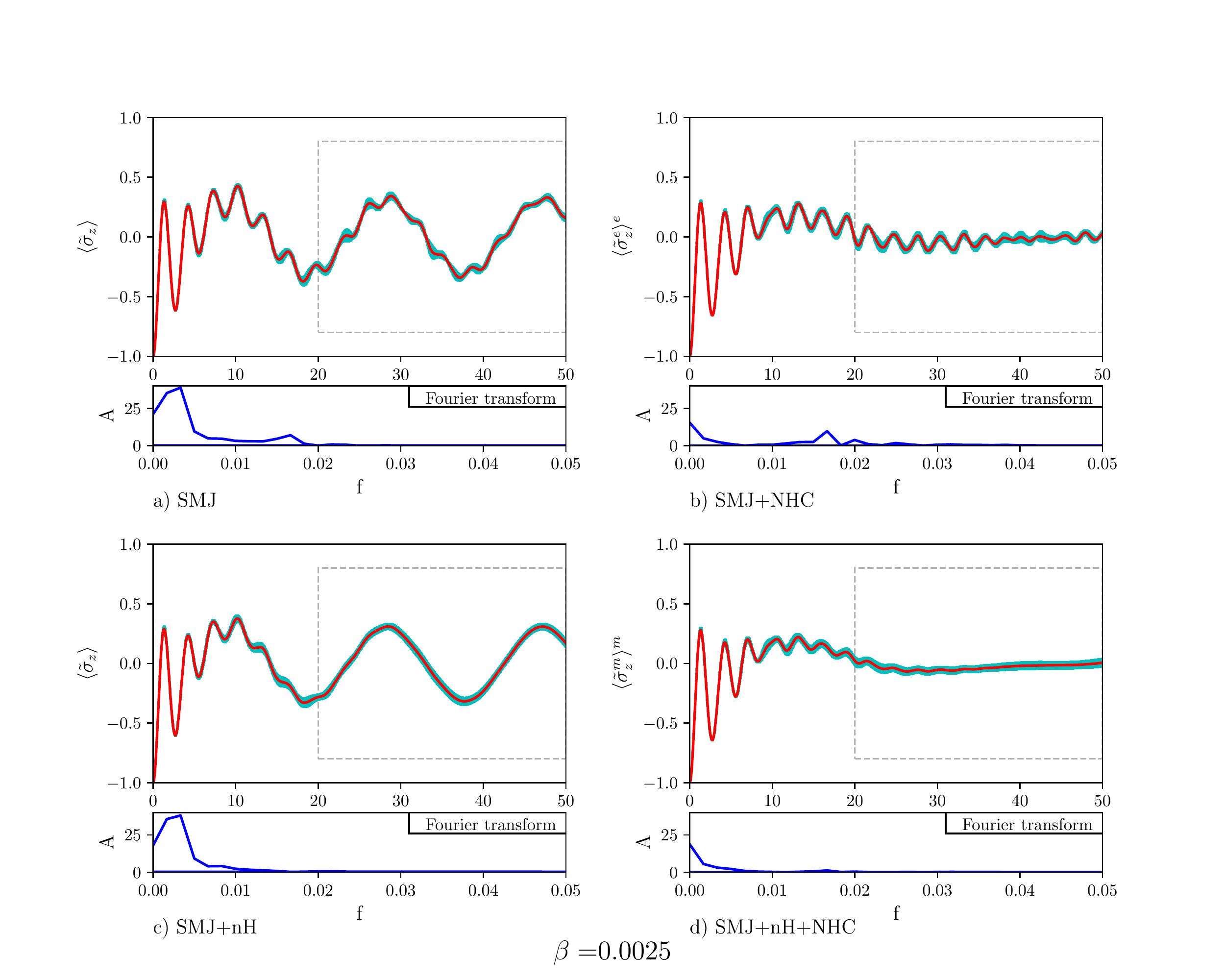}
\caption{The plot shows the average population transfer
vs. time for different types of dynamics.
NMDS=10000, NMCP=2500, $\tau=0.005$, $\gamma = 0.1$, $\omega = 1/3$,
$\Delta = 1$, $g = 0.007$, $\beta = 0.0025$. The red line is
the average trend, the cyan area is the error.
The blue line is the Fourier transform, carried out after
the stabilization of the trend (from $t=20$ onward).
In Panel (a) the population difference oscillates with
low and high frequency contributions (SMJ).
Panel (b) shows that the
thermostat (SMJ+NHC) facilitates the population transfer
and stabilizes it more than in the case of lower temperatures.
The low frequency terms are highly damped while the high frequency
ones are unaltered. Panel (c) shows that the effect of
the decay operator (SMJ+nH) is to cancel the high frequency term.
Finally, Panel (d) shows that SMJ+nH+NHC
displays a stable transfer:
both the low and the high frequency terms are strongly damped.
} 
\label{FIG7:SigmaTot_0.0025}
\end{figure}
%%%%%%%%%%%%%%%%%%%%%%%%%%%%%%%%%%%

\noindent
{\bf Funding:} This research received no external funding.

\noindent
{\bf Acknowledgments:} AM and AS express their thanks to Hiromichi Nakazato for carefully reading 
the manuscript and for numerous stimulating suggestions.
All the authors are grateful to Dr. Marika Matalone for drawing 
Figure \ref{Fig1:ModelScheme}.

\appendix

\section{Time evolution of the trace of $\tilde\Omega(t)$}
\label{app:qce-nherm}

The trace of Eq. (\ref{eq:ddt-Omega_matrix_W}):
\ba
\frac{d}{dt}\tilde{\rm T}{\rm r}\left(\tilde\Omega(t)\right)
&=&
- 2\tilde{\rm T}{\rm r}\left(\hat\Gamma\tilde\Omega(X,t)\right)
+\frac{1}{2}\tilde{\rm T}{\rm r}\left(
\left\{\tilde{H}^{\rm e},\tilde\Omega(t)\right\}
_{\mbox{\boldmath$\cal B$}} \right) \nonumber\\
&-&
\frac{1}{2}\tilde{\rm T}{\rm r}\left(
\left\{\tilde\Omega(t),\tilde{H}\right\}_{\mbox{\boldmath$\cal B$}} \right)
\;.
\label{eq:dotxi}
\ea
The last term in the right hand side of Eq. (\ref{eq:dotxi}) can be transformed
as follows
\ba
-\frac{1}{2}\tilde{\rm T}{\rm r}\left(
\left\{\tilde\Omega(t),\tilde{H}\right\}_{\mbox{\boldmath$\cal B$}} \right)
&=&
-\frac{1}{2}\tilde{\rm T}{\rm r}\left(\sum_{I,J=1}^{2N}
\frac{\partial\tilde\Omega(t)}{\partial X_I}{\cal B}_{IJ}
\frac{\partial\tilde{H}}{\partial X_J}\right)
=
-\frac{1}{2}\tilde{\rm T}{\rm r}\left(\sum_{I,J=1}^{2N}
\frac{\partial\tilde\Omega(t)}{\partial X_J}{\cal B}_{JI}
\frac{\partial\tilde{H}}{\partial X_I}\right)
\nonumber\\
&=&
\frac{1}{2}\tilde{\rm T}{\rm r}\left(\sum_{I,J=1}^{2N}
\frac{\partial\tilde{H}}{\partial X_I} {\cal B}_{IJ}
\frac{\partial\tilde\Omega(t)}{\partial X_J}\right)
=\frac{1}{2}\tilde{\rm T}{\rm r}\left(\left\{
\tilde{H},\tilde\Omega(t)\right\}_{\mbox{\boldmath$\cal B$}}\right)
\label{eq:Tr_Poisson}
\ea
Substituting Eq. (\ref{eq:Tr_Poisson}) in the right hand side of Eq.
(\ref{eq:dotxi}), one obtains
\ba
\frac{d}{dt}\tilde{\rm T}{\rm r}\left(\tilde\Omega(t)\right)
&=&
- 2\tilde{\rm T}{\rm r}\left(\hat\Gamma\tilde\Omega(X,t)\right)
+\tilde{\rm T}{\rm r}\left(
\left\{\tilde{H},\tilde\Omega(t)\right\}
_{\mbox{\boldmath$\cal B$}} \right)
\label{eq:dotxi_2} \;.
\ea
However, the last term in the right hand side of Eq. (\ref{eq:dotxi_2})
is zero:
\ba
\tilde{\rm T}{\rm r}\left(
\left\{\tilde{H},\tilde\Omega(t)\right\}
_{\mbox{\boldmath$\cal B$}} \right)
&=&
\tilde{\rm T}{\rm r}\left(\sum_{I,J=1}^{2N}
\frac{\partial\tilde{H}}{\partial X_I} {\cal B}_{IJ}
\frac{\partial\tilde\Omega(t)}{\partial X_J} \right)
=- \tilde{\rm T}{\rm r}\left(\sum_{I,J=1}^{2N}
\frac{\partial^2\tilde{H}}{\partial X_I\partial X_J} {\cal B}_{IJ}
\tilde\Omega(t)\right)
\label{eq:corr_Poisson}
\nonumber\\
&=&0
\ea
because of the antisymmetry of $\mbox{\boldmath$\cal B$}$.
Equation (\ref{eq:corr_Poisson}) allows one
to obtain Eq. (\ref{eq:ddt-qcTraceOm}).

%%%%%%%%%%%%%%%%%%%%%%%%%%%%%%%%%%%%%%%%%%%%%%%%%%%%%%%%%%%%%%%%%%%%%%%

\section{Time evolution of the trace of $\tilde{\Omega}^{\rm e}(t)$}
\label{app:qce-nherm_nhc}

The trace of Eq. (\ref{eq:ddt-Omega_e_matrix}) is
\be
\frac{d}{dt}\tilde{\rm T}{\rm r}^{\rm e}\left(\tilde{\Omega}^{\rm e}(t)\right)
=
-2\tilde{\rm T}{\rm r}^{\rm e}\left(\hat\Gamma\tilde{\Omega}^{\rm e}(t)\right)
+\frac{1}{2}\tilde{\rm T}{\rm r}^{\rm e}
\left(\left\{\tilde{H}^{\rm e},\tilde\Omega^{\rm e}
\right\}_{\mbox{\boldmath$\cal B$}^{\rm e}}\right)
-\frac{1}{2}\tilde{\rm T}{\rm r}^{\rm e}
\left(\left\{ \tilde\Omega^{\rm e},\tilde{H}^{\rm e}
\right\}_{\mbox{\boldmath$\cal B$}^{\rm e}}\right)
-\tilde{\rm T}{\rm r}^{\rm e}\left(\kappa^{\rm e}\tilde{\Omega}^{\rm e}(t)\right)
\;.
\label{eq:ddt-Tr-Omega_e_matrix}
\end{equation}
One can find the following identity
\ba
\tilde{\rm T}{\rm r}^{\rm e}
\left(\left\{ \tilde\Omega^{\rm e},\tilde{H}^{\rm e}
\right\}_{\mbox{\boldmath$\cal B$}^{\rm e}}\right)
&=&
\tilde{\rm T}{\rm r}^{\rm e}\left(
\sum_{I,J=1}^{2(N+2)}\frac{\partial \tilde\Omega^{\rm e}}{\partial X_I^{\rm e}}
{\cal B}_{IJ}^{\rm e}\frac{\partial\tilde{H}^{\rm e}}{\partial X_J^{\rm e}}
\right)
=
\tilde{\rm T}{\rm r}^{\rm e}\left(
\sum_{I,J=1}^{2(N+2)}\frac{\partial \tilde\Omega^{\rm e}}{\partial X_J^{\rm e}}
{\cal B}_{JI}^{\rm e}\frac{\partial\tilde{H}^{\rm e}}{\partial X_I^{\rm e}}
\right)
\nonumber\\
&=&
\tilde{\rm T}{\rm r}^{\rm e}\left(
\sum_{I,J=1}^{2(N+2)} \frac{\partial\tilde{H}^{\rm e}}{\partial X_I^{\rm e}}
{\cal B}_{JI}^{\rm e} \frac{\partial \tilde\Omega^{\rm e}}{\partial X_J^{\rm e}}
\right)
=- \tilde{\rm T}{\rm r}^{\rm e}\left(
\sum_{I,J=1}^{2(N+2)} \frac{\partial\tilde{H}^{\rm e}}{\partial X_I^{\rm e}}
{\cal B}_{IJ}^{\rm e} \frac{\partial \tilde\Omega^{\rm e}}{\partial X_J^{\rm e}}
\right)
\nonumber\\
&=&
-\tilde{\rm T}{\rm r}^{\rm e}
\left(\left\{ \tilde{H}^{\rm e} , \tilde\Omega^{\rm e}
\right\}_{\mbox{\boldmath$\cal B$}^{\rm e}}\right)
\label{eq:Tre-Oe-He}
\ea
Inserting Eq. (\ref{eq:Tre-Oe-He}) in Eq. (\ref{eq:ddt-Tr-Omega_e_matrix})
one obtains
\be
\frac{d}{dt}\tilde{\rm T}{\rm r}^{\rm e}\left(\tilde{\Omega}^{\rm e}(t)\right)
=
-2\tilde{\rm T}{\rm r}^{\rm e}\left(\hat\Gamma\tilde{\Omega}^{\rm e}(t)\right)
+\tilde{\rm T}{\rm r}^{\rm e}
\left(\left\{\tilde{H}^{\rm e},\tilde\Omega^{\rm e}
\right\}_{\mbox{\boldmath$\cal B$}^{\rm e}}\right)
-\tilde{\rm T}{\rm r}^{\rm e}\left(\kappa^{\rm e}\tilde{\Omega}^{\rm e}(t)\right)
\;.
\label{eq:ddt-Tr-Omega_e_matrix_2}
\ee
One can now find the following identity:
\ba
\tilde{\rm T}{\rm r}^{\rm e}
\left(\left\{\tilde{H}^{\rm e},\tilde\Omega^{\rm e}
\right\}_{\mbox{\boldmath$\cal B$}^{\rm e}}\right)
&=&
-\tilde{\rm T}{\rm r}^{\rm e} \left(\left\{\tilde{H}^{\rm e},\tilde\Omega^{\rm e}
\right\}_{\mbox{\boldmath$\cal B$}^{\rm e}}\right)
=
-\tilde{\rm T}{\rm r}^{\rm e}\left(\sum_{I,J=1}^{2(N+1)}
\frac{\partial\tilde{H}^{\rm e}}{\partial X_I^{\rm e}}
{\cal B}_{IJ}^{\rm e}
\frac{\partial\tilde\Omega^{\rm e}}{\partial X_J^{\rm e}}
\right)
\nonumber\\
&=&
\tilde{\rm T}{\rm r}^{\rm e}\left(\sum_{I,J=1}^{2(N+1)}
\frac{\partial {\cal B}_{IJ}^{\rm e} }{\partial X_J^{\rm e}}
\frac{\partial\tilde{H}^{\rm e}}{\partial X_I^{\rm e}}
\tilde\Omega^{\rm e} \right)
=
\tilde{\rm T}{\rm r}^{\rm e}\left(\kappa^{\rm e} \tilde\Omega^{\rm e} \right) \;.
\label{eq:Tre-ke-Oe}
\ea
Equation (\ref{eq:Tre-ke-Oe}) is obtained exploiting the identity
\be
\sum_{I,J=1}^{2(N+1)}{\cal B}_{IJ}^{\rm e}
\frac{\partial^2\tilde{H}^{\rm e}}{\partial X_I^{\rm e}\partial X_J^{\rm e}}
\equiv 0 \;,
\ee
which derives from the fact that the trace of a symmetric matrix times
an antisymmetric matrix is identically zero.
Finally, substituting Eq. (\ref{eq:Tre-ke-Oe}) in Eq. (\ref{eq:ddt-Tr-Omega_e_matrix_2}),
one obtains Eq. (\ref{eq:ddt-qcTraceOm_e}).

%%%%%%%%%%%%%%%%%%%%%%%%%%%%%%%%%%%%%%%%%%

\end{document}